\documentclass[12pt]{article}
\usepackage{graphicx} % Required for inserting images
\usepackage{subcaption}
\usepackage{amsmath,amssymb,xcolor,physics}
\usepackage{url}
\usepackage{ulem}
\usepackage{hyperref}
\usepackage{soul}  % for strikethrough text \st
\usepackage{tikz}
\usepackage{pgfplots}
\usepackage{cite}
\usetikzlibrary{fadings}
\pgfplotsset{compat=1.18}

\hypersetup{
    colorlinks=true,   % Set to false for boxed links
    linkcolor=blue,    % Color for internal links
    citecolor=red,     % Color for citations
    filecolor=magenta, % Color for file links
    urlcolor=cyan      % Color for URLs
}

\allowdisplaybreaks[4]

\begin{document}

\begin{center}
{\Large Universal framework with exponential speedup}\\
\vspace{3mm}
{\Large for the quantum simulation of quantum field theories}\\
\vspace{3mm}
{\Large including QCD}

\end{center}
\vspace{0.1cm}
\vspace{0.1cm}
\begin{center}
Jad C.~Halimeh,$^{\rm Q}$ Masanori Hanada,$^{\rm C}$ and Shunji Matsuura$^{\rm D}$
\end{center}
\vspace{0.3cm}

\begin{center}
$^{\rm Q}$\, Max Planck Institute of Quantum Optics, 85748 Garching, Germany\\
\vspace{1mm}
$^{\rm Q}$\, Department of Physics and Arnold Sommerfeld Center for Theoretical Physics, Ludwig Maximilian University of Munich, 80333 Munich, Germany\\
\vspace{1mm}
$^{\rm Q}$\, Munich Center for Quantum Science and Technology, 80799 Munich, Germany\\
\vspace{1mm}
$^{\rm C}$\, School of Mathematical Sciences, Queen Mary University of London\\
Mile End Road, London, E1 4NS, United Kingdom\\
\vspace{1mm}
$^{\rm C}$\, qBraid Co., Harper Court 5235, Chicago, IL 60615, United States\\
\vspace{1mm}
$^{\rm D}$\, RIKEN Center for Interdisciplinary Theoretical and Mathematical Sciences(iTHEMS)\\
RIKEN, Wako, Saitama 351-0198, Japan\\
\vspace{1mm}
$^{\rm D}$\, Department of Electrical and Computer Engineering,
University of British Columbia
\\
Vancouver, BC V6T 1Z4, Canada\\
\vspace{1mm}
$^{\rm D}$\, Department of Physics, University of Guelph, ON N1G 1Y2, Canada \\
\vspace{1mm}
$^{\rm D}$\, Center for Mathematical Science and Advanced Technology\\
Japan Agency for Marine-Earth Science and Technology,
Yokohama 236-0001, Japan 
\end{center}

\vspace{0.5cm}

\begin{center}
  {\bf Abstract}
\end{center}
We present a quantum simulation framework universally applicable to a wide class of quantum systems, including quantum field theories such as quantum chromodynamics (QCD). 
Specifically, we generalize an efficient quantum simulation protocol developed for bosonic theories in Ref.~\cite{Halimeh:2024bth} which, when applied to Yang--Mills theory, demonstrated an exponential resource advantage with respect to the truncation level of the bosonic modes, to systems with both bosons and fermions using the Jordan--Wigner transform and also the Verstraete--Cirac transform. We apply this framework to QCD using the orbifold lattice formulation and achieve an exponential speedup compared to previous proposals. As a by-product, exponential speedup is achieved in the quantum simulation of the Kogut--Susskind Hamiltonian, the latter being a special limit of the orbifold lattice Hamiltonian. In the case of Hamiltonian time evolution of a theory on an $L^d$ spatial lattice via Trotterization, one Trotter step can be realized using $\mathcal{O}(L^d)$ numbers of CNOT gates, Hadamard gates, phase gates, and one-qubit rotations. We show this analytically for any matter content and $\mathrm{SU}(N)$ gauge group with any $N$. Even when we use the Jordan--Wigner transform, we can utilize the cancellation of quantum gates to significantly simplify the quantum circuit. We also discuss a block encoding of the Hamiltonian as a linear combination of unitaries using the Verstraete--Cirac transform. Our protocols do not assume oracles, but rather present explicit constructions with rigorous resource estimations without a hidden cost, and are thus readily implementable on a quantum computer.

\newpage
\tableofcontents
%%%%%%%%%%%%%%%%%%%%%%
%%%%%%%%%%%%%%%%%%%%%%
\section{Introduction}
%%%%%%%%%%%%%%%%%%%%%%
%%%%%%%%%%%%%%%%%%%%%%
The quantum simulation of high-energy physics is currently an extremely active research area establishing connections to other fields such as quantum many-body dynamics, condensed matter, quantum information theory, and quantum computing~\cite{Dalmonte:2016alw,Banuls:2019bmf, Zohar_review, Aidelsburger:2021mia, Zohar_NewReview, klco2021standard, Bauer:2022hpo, DiMeglio:2023nsa, Cheng_review, Halimeh_review, Cohen:2021imf, Lee:2024jnt, Turro:2024pxu}. Recent years have seen an explosion in theoretical proposals \cite{Zohar:2011cw,Zohar:2012ay,Banerjee2012,Zohar2013,Banerjee2013atomic,Halimeh2022tuning,Cheng2022,Fontana2022,surace2023abinitio,osborne2022largescale,osborne2023spins} and experimental realizations \cite{Martinez2016,Klco2018,Goerg2019,Schweizer2019,Mil2020,Yang2020,Wang2021,Zhou2022,Wang2023,Zhang:2023hzr,Ciavarella2024quantum,Ciavarella:2024lsp,de2024observationstringbreakingdynamicsquantum,liu2024stringbreakingmechanismlattice,Farrell:2023fgd,Farrell:2024fit,zhu2024probingfalsevacuumdecay,Ciavarella:2021nmj,Ciavarella:2023mfc,Ciavarella:2021lel,Gustafson:2023kvd,Gustafson:2024kym,Lamm:2024jnl,Farrell:2022wyt,Farrell:2022vyh,Li:2024lrl,Zemlevskiy:2024vxt,Lewis:2019wfx,Atas:2021ext,ARahman:2022tkr,Atas:2022dqm,Mendicelli:2022ntz,Kavaki:2024ijd,Than:2024zaj,Angelides2025first,alexandrou2025realizingstringbreakingdynamics,Cochran:2024rwe,Gyawali:2024hrz,gonzalezcuadra2024observationstringbreaking2,crippa2024analysisconfinementstring2,schuhmacher2025observationhadronscatteringlattice,davoudi2025quantumcomputationhadronscattering} on analog and digital quantum simulation platforms ranging from neutral atoms to trapped ions and superconducting qubits, highlighting the tremendous interest in this relatively young research frontier.
Nevertheless, most of this body of work has focused on Abelian lattice gauge theories in one or two spatial dimensions. Despite the fascinating phenomenology uncovered by the above quantum simulation experiments of lattice gauge theories, including discovering novel quantum many-body dynamics such as scarring \cite{Bernien2017,Su2022,Surace2020,Desaules2021,Desaules2022prominent,Banerjee2021,Halimeh2022robust,biswas2022scars,Sau:2023clm,osborne2024quantummanybodyscarring21d,Budde2024qmbs,Calajo2025QMBS}, Hilbert-space fragmentation \cite{Sala2020ergodicity,Khemani2020localization,Jeyaretnam2025Hilbert,ciavarella2025generichilbertspacefragmentation,datla2025statisticallocalizationrydbergsimulator}, and disorder-free localization \cite{Smith2017,Brenes2018,Gyawali:2024hrz}, the overarching goal of this field --- the implementation of $3+1$D quantum chromodynamics (QCD) on quantum hardware and extraction of relevant physics from quantum simulations~\cite{Bauer:2022hpo,DiMeglio:2023nsa} --- remains an open problem. In this paper, we focus on digital quantum simulation and provide a concrete strategy for QCD in $3+1$ dimensions. 

In this vein, the challenge lies in formulating the gauge theory with fermionic matter on a spatial lattice, which can then be implemented on a quantum device. The gauge field has an infinite-dimensional local Hilbert space that, in general, requires truncation to be implemented on a digital quantum computer. However, technical difficulties associated with the \underline{compact variables} (unitary link variables) have hampered attempts to implement non-Abelian gauge theories, particularly with fermionic matter in more than one spatial dimension. For one spatial dimension, the gauge degrees of freedom can be integrated out \cite{Muschik2017}. However, this leads to long-range interactions between matter fields, restricting any realistic experimental implementation to a few sites \cite{Martinez2016,Atas:2021ext}. Alternatively, one can also employ Gauss's law to integrate out the matter fields \cite{Surace2020,Desaules2021,Desaules2022prominent}, but this is restricted to bosonic matter degrees of freedom and can lead to long-range gates for the minimal coupling in spatial dimensions greater than one~\cite{Joshi2025efficient}.

The Kogut--Susskind formulation of lattice gauge theories in its original form with compact variables has been the standard approach for quantum simulation efforts. This formulation works well for Abelian gauge theories or systems in one spatial dimension. However, extending quantum simulation to non-Abelian gauge theories in higher spatial dimensions has proven challenging. Despite sustained efforts over two decades since the pioneering work of Ref.~\cite{Byrnes:2005qx}, a systematic and scalable simulation protocol remains elusive.\footnote{Several theoretical proposals have been developed with varying degrees of implementational complexity \cite{Klco2020,Raychowdhury2020loop,Halimeh2023Spin,Surace2023scalable,Fontana:2024rux,illa2025improvedhoneycombhyperhoneycomblattice,depaciani2025quantumsimulationfermionicnonabelian}.}
A fundamental bottleneck emerges at the stage of programming the Kogut--Susskind Hamiltonian on quantum devices~\cite{Hanada:2025yzx}. The truncated Hamiltonian requires extensive classical preprocessing to design implementable quantum circuits in terms of native gates. Unfortunately, the computational cost of this classical preprocessing appears to scale exponentially, presenting a significant obstacle. While it might seem feasible to design efficient circuits through direct construction, this approach has not yielded scalable solutions despite considerable community effort. This creates a paradoxical situation: even with the advent of fault-tolerant quantum computers possessing billions of logical qubits, realizing quantum advantage would remain hindered by the classical intractability of circuit preparation.
Recent work has addressed this challenge by reformulating the Kogut--Susskind Hamiltonian as a special limit of the orbifold lattice Hamiltonian \cite{bergner2025exponentialspeedupquantumsimulation}, thereby rendering circuit design an easy exercise.

The orbifold lattice Hamiltonian~\cite{Buser:2020cvn,Bergner:2024qjl,Kaplan:2002wv} allows us to formulate non-Abelian gauge theories on a lattice in a drastically simple manner: because the orbifold lattice Hamiltonian is formulated using \underline{noncompact variables}, \textit{all problems associated with compact variables are absent by construction}. Most importantly, no group theory is needed at all. 
The use of the noncompact variables amounts to introducing scalar fields that decouple from low-energy physics. One can even obtain the Kogut--Susskind Hamiltonian as a special limit of the orbifold lattice Hamiltonian, and hence the same physics --- including lattice artifacts --- can be studied using the orbifold lattice Hamiltonian \cite{bergner2025exponentialspeedupquantumsimulation}. 
The use of noncompact variables makes it straightforward to build efficient algorithms for any SU($N$), any matter content, and any spatial dimensions in a unified manner. Specifically, Yang--Mills theory defined on the orbifold lattice is expressed as a special case of a generic class of Hamiltonians, 
\begin{align}
\hat{H}_{\rm bos}
=
\frac{1}{2}\sum_a\hat{p}_a^2
+
V(\hat{x})\, , 
\label{eq:Hamiltonian-boson}
\end{align}
with the canonical commutation relations
\begin{align}
[\hat{x}_a,\hat{p}_b]
=
\mathrm{i}\delta_{ab}\, , 
\qquad
[\hat{x}_a,\hat{x}_b]
=
[\hat{p}_a,\hat{p}_b]
=
0\, . 
\end{align}
Ref.~\cite{Halimeh:2024bth} provided a universal framework for the quantum simulation of this class that achieves exponential speedup with regard to the number of qubits assigned to each bosonic mode. For lattice gauge theory, this is intimately associated with the speedup with regard to the lattice spacing, because more qubits per boson are required as one approaches the continuum limit~\cite{Hanada:2025yzx}. No oracle is needed, and all steps are expressed explicitly without a black box. To the best of our knowledge, this universal framework is the only programmable approach to quantum simulation of Yang--Mills theory for $N\ge 2$ and three spatial dimensions.\footnote{
There has been a proposal~\cite{Rhodes:2024zbr} for a potential exponential speedup based on the vanilla Kogut--Susskind Hamiltonian, which could be implemented if certain oracles were realized efficiently using native quantum gates on a quantum computer. Explicit construction of such oracles is not known, and finding efficient circuit implementations of such oracles may well be an exponentially hard problem. 
}

In this paper, we will generalize this framework to systems consisting of bosons and fermions, which is particularly pertinent to the current drive to realize quantum simulation experiments of lattice gauge theories in more than one spatial dimension~\cite{Cochran:2024rwe,Gyawali:2024hrz,gonzalezcuadra2024observationstringbreaking2,crippa2024analysisconfinementstring2}, where fermionic matter can lead to distinct physics \cite{srivatsa2025bosonicvsfermionicmatter}. Specifically, we consider the class of Hamiltonians
\begin{align}
\hat{H}
=
\hat{H}_{\rm bos}
+
V_{\rm fer}(\hat{x},\hat{\psi})\, , 
\label{eq:Hamiltonian-boson+fermion}
\end{align}
where fermions $\hat{\psi}_1,\hat{\psi}_2,\ldots$ are real and satisfy the canonical (anti)commutation relations 
\begin{align}
\big\{\hat{\psi}_i,\hat{\psi}_j^\dagger\big\}
=
2\delta_{ij}, 
\qquad
\big[\hat{x}_a,\hat{\psi}_i\big]
=
\big[\hat{p}_a,\hat{\psi}_i\big]
=
0\, . 
\end{align}
This is a rather generic class that contains orbifold lattice QCD~\cite{Bergner:2024qjl}.

For simplicity, we assume $V_{\rm fer}(\hat{x},\hat{\psi})$ to be a polynomial.
The orbifold lattice Hamiltonian for QCD is in this class. This is not too restrictive because, even if it is not, we can truncate the Taylor expansion of $V_{\rm fer}$ and utilize a polynomial to approximate the full potential. 

Unlike bosons, fermions do not require truncation. The number of logical qubits needed to describe fermions is the same as the number of fermions.\footnote{
More precisely, we count the number of complex fermions. Furthermore, we count different components of representations concerning the gauge group or spinors independently.} 
However, the Hamiltonian can become nonlocal in terms of qubits. This is because long Pauli strings are needed to reproduce the anti-commuting nature of fermions. This makes the quantum circuit complicated. With a naive implementation, the simulation cost can grow faster than the volume, which is a significant slowdown compared to the universal simulation protocol for bosonic theories. (Exponential speedup compared to the classical method is still achieved, though.) In this paper, we show that this problem can be resolved by using a few simple tricks. In Sec.~\ref{sec:universal_framework_boson+fermion}, we apply the Jordan--Wigner transform, which is the simplest, and arguably a naive, implementation, to the Hamiltonian time evolution via Trotter decomposition. Long Pauli strings can be converted to products of many CNOT gates, which mostly cancel with each other when the ordering of operators is chosen carefully. This method is suitable for the NISQ devices. In Sec.~\ref{sec:Verstraete-Cirac}, we use the Verstraete--Cirac transform, which eliminates long Pauli strings by cleverly introducing additional fermions. This approach enables us to use not just the Trotter decomposition but also the block encoding of a linear combination of unitaries, which can be a powerful tool on fault-tolerant quantum computers. Needless to say, Jordan--Wigner and Verstraete--Cirac transforms are far from the only options. We will briefly mention other protocols, too. 

This paper is organized as follows. Sec.~\ref{sec:universal_framework_boson} is a review of the universal framework for bosonic systems~\cite{Halimeh:2024bth}. We provide only minimal materials that are necessary to understand the generalization to include fermions. Specifically, the readers should find the details of the orbifold lattice Hamiltonian in other papers, e.g., Refs.~\cite{Bergner:2024qjl}. (Because of the universal nature of our approach, we do not even need the details of the Hamiltonian to understand and generalize it.) 
In the later sections, fermions are added to this framework. As a concrete example, QCD is considered. In Sec.~\ref{sec:universal_framework_boson+fermion}, we investigate the Jordan--Wigner transform. It turns out that Jordan--Wigner transform is already rather powerful when combined with the universal framework and applied to the Hamiltonian time evolution via Trotter decomposition. In Sec.~\ref{sec:Verstraete-Cirac}, we apply the Verstraete--Cirac transform and see that the Trotter decomposition and the block encoding of the Hamiltonian as a linear combination of unitaries are straightforward. We comment on other possibilities in Sec.~\ref{sec:other_methods}. Sec.~\ref{sec:conclusion} provides the conclusion and discussions. 

%%%%%%%%%%%%%%%%%%%%%%
%%%%%%%%%%%%%%%%%%%%%%
\section{Universal framework for bosonic systems: a review
}\label{sec:universal_framework_boson}
%%%%%%%%%%%%%%%%%%%%%%
%%%%%%%%%%%%%%%%%%%%%%
We now provide a brief review of the universal framework for quantum simulations of bosonic systems~\cite{Halimeh:2024bth}. As a concrete algorithm, we discuss the Trotter decomposition of Hamiltonian time evolution. Block encoding as a linear combination of unitaries (LCU) in this framework has been discussed in Ref.~\cite{Hanada:2025yzx}. 

%%%%%%%%%%%%%%%%%%%%%%%%%%%
%%%%%%%%%%%%%%%%%%%%%%%%%%%
\subsection{Truncation of Hilbert space and operators}
%%%%%%%%%%%%%%%%%%%%%%%%%%%
%%%%%%%%%%%%%%%%%%%%%%%%%%%

%%%%%%%%%%%%%%%%%%%%%%%%%%%
%%%%%%%%%%%%%%%%%%%%%%%%%%%
\subsubsection{Coordinate basis and $\hat{x}$}
%%%%%%%%%%%%%%%%%%%%%%%%%%%
%%%%%%%%%%%%%%%%%%%%%%%%%%%
Let us employ $\vec{x}$ to represent the coordinates of all $N_b$ bosons collectively.
When utilizing this notation, we indicate that the coordinate eigenstate of the system takes the form 
\begin{align}
\ket{\vec{x}}
=
\otimes_a\ket{x_a}\, ,
\end{align}
where each individual boson possesses a coordinate eigenstate $\ket{x_a}$ ($a=1,2,\cdots, N_b$).
The coordinate eigenstate $\ket{\vec{x}}$ satisfies the relation 
\begin{align}
\hat{\vec{x}}\ket{\vec{x}}
=
\vec{x}\ket{\vec{x}}\, . 
\end{align}
Additionally, we view the Hilbert space as a tensor product comprising the Hilbert spaces of the separate bosons in the coordinate eigenbasis:
\begin{align}
\mathcal{H}
=
\otimes_a\mathcal{H}_a\, , 
\qquad
\mathcal{H}_a
=
\mathrm{Span}\{\ket{x_a}|x_a\in\mathbb{R}\}\, . 
\end{align}
To this point, every boson has a wavefunction expressible in the basis given by $\ket{x_a}$, residing in an infinite-dimensional Hilbert space.
As an illustration, one might consider a one-dimensional quantum oscillator existing in a superposition of various locations.
To transform the problem into a finite-dimensional Hilbert space, we introduce a cutoff for each boson coordinate $x_a$. 
We allocate $Q$ qubits to each boson. Each $x_a$ may assume $\Lambda=2^Q$ distinct values. 
For practical purposes, it proves advantageous to implement periodic boundary conditions $x+2R\sim x$. 
In this framework, a suitable notation is
\begin{align}
   \delta_x=\frac{2R}{\Lambda} 
\end{align}
and 
\begin{align}
    x_a
    =
    \pm\frac{\delta_x}{2},\, \pm\frac{3\delta_x}{2}\, \cdots\, \pm\frac{(\Lambda-1)\delta_x}{2}\, . 
\end{align}  
We need to take $R$ sufficiently large so that $x\sim\pm R$ states are not significantly excited, ensuring that the boundary condition does not influence the relevant physics. 
Analysis of truncation effects typically depends on the specific system under investigation, including its parameters such as the coupling constant.
Often, these systematic effects require numerical investigation. A study focusing on expectation values calculated via classical sampling methods was used in Ref.~\cite{Hanada:2022pps}.
Using Pauli $Z$ operators defined as  
\begin{align}
    Z \equiv \ket{0}\bra{0} - \ket{1}\bra{1}\, ,  
    \label{eq: pauli Z}
\end{align}
we can express $\hat{x}_a$ as  
\begin{align}
\hat{x}_a
=
-
\delta_x\cdot
\left(
\frac{Z_{a,1}}{2}
+
2\cdot\frac{Z_{a,2}}{2}
+
\cdots
+
2^{Q-1}\cdot\frac{Z_{a,Q}}{2}
\right)\, . 
\end{align}
Here, $Z_{a,j}$ represents the Pauli $Z$ operator acting on the $j$-th qubit assigned to the $a$-th boson. 
Let $V$ be a polynomial of degree $n$. Then, numerous $n$-boson couplings exist in the form $\hat{x}_{a_1}\otimes\cdots\otimes\hat{x}_{a_n}$. 
Each $\hat{x}$ comprises a weighted sum of $Z_{1}$,..., $Z_{Q}$. 
Therefore, each $n$-boson coupling contains $Q^n$ terms, with each term being a tensor product of at most four $Z$'s. 
%%%%%%%%%%%%%%%%%%%%%%%%%%%
%%%%%%%%%%%%%%%%%%%%%%%%%%%
\subsubsection{Momentum basis and $\hat{p}$}\label{sec:Fourier_transform}
%%%%%%%%%%%%%%%%%%%%%%%%%%%
%%%%%%%%%%%%%%%%%%%%%%%%%%%
We implement the Fourier transform to the momentum basis:
\begin{align}
    \ket{p_a}
    =
    \frac{1}{\sqrt{\Lambda}}\sum_{x_a} e^{2\pi \mathrm{i}p_ax_a}\ket{x_a}\, . 
\end{align}
Here, $p_a$ takes values $\pm\frac{\delta_p}{2}$, $\pm\frac{3\delta_p}{2}$, ..., $\pm\frac{(\Lambda-1)\delta_p}{2}$, where 
\begin{align}
\delta_p
=
\frac{\pi}{R}\, . 
\end{align}
Then, within the momentum basis, we can take
\begin{align}
\hat{p}_a
=
-
\delta_p\cdot
\left(
\frac{Z_{a,1}}{2}
+
2\cdot\frac{Z_{a,2}}{2}
+
\cdots
+
2^{Q-1}\cdot\frac{Z_{a,Q}}{2}
\right)\, .  
\end{align}
An important feature is that the Fourier transformation can be executed for all bosons simultaneously in parallel, meaning the circuit depth is determined solely by the truncation level $\Lambda$ and is \emph{independent} of the total number of bosons. Furthermore, we can use one of the most powerful quantum algorithms: quantum Fourier transform~\cite{Coppersmith:2002skh}. This is a significant advantage over the vanilla Kogut--Susskind approach, which lacks efficient Fourier transformation algorithms. 

%%%%%%%%%%%%%%%%%%%%%%%%%%%
%%%%%%%%%%%%%%%%%%%%%%%%%%%
\subsection{Hamiltonian time evolution via Trotterization}\label{sec:Trotter_cost_piece}
%%%%%%%%%%%%%%%%%%%%%%%%%%%
%%%%%%%%%%%%%%%%%%%%%%%%%%%
For simplicity, suppose that $V$ is a polynomial of degree $n$. Generalization to a more generic case is straightforward. 

We can consider the momentum part and the interaction part separately. 
Because $[\hat{p}_j,\hat{p}_k]=0$, the momentum part factorizes as 
\begin{align}
\exp\left(-\textrm{i}\theta\sum_j\hat{p}^2_j\right)
=
\prod_{j}\exp\left(-\textrm{i}\theta\hat{p}^2_j\right)\, . 
\end{align}
Therefore, the total cost is that of one boson times the number of bosons.
The potential part factorizes as well.
Schematically, it takes the form 
\begin{align}
\prod_{j_1,\cdots,j_n}\exp\left(-\textrm{i}\theta \mathcal{C}_{j_1\cdots j_n}\, 
\hat{x}_{j_1}\cdots\hat{x}_{j_n}\right)\, , 
\end{align}
with certain coefficients $\mathcal{C}_{j_1\cdots j_n}$. 

As mentioned before, $\hat{x}_{j_1}\cdots\hat{x}_{j_n}$ is a weighted sum of tensor products of $n$ Pauli $Z$'s. 
Schematically, the interaction term is a product of the Pauli rotations, 
\begin{align}
\prod_{a_1\cdots a_n}\exp\left(-\textrm{i}\theta \mathcal{C}'_{a_1\cdots a_n}Z_{a_1}\cdots Z_{a_n}\right)\, . 
\label{eq:interaction-generic-Z}
\end{align}
In the same way, in the momentum basis, we can write $e^{-\mathrm{i}\theta\hat{p}^2}$ as
\begin{align}
    \exp\left(-\mathrm{i}\theta\hat{p}^2\right)
    =
    \prod_{a<b}
    \exp\left(-\mathrm{i}\theta \mathcal{C}''_{ab}Z_{a}Z_{b}\right)\, . 
    \label{eq: diagonal kinetic terms}
\end{align}

These Pauli rotations can be implemented through CNOT gates combined with single-qubit rotations. 
The CNOT gate with the control qubit $a$ and target qubit $b$ is defined by 
\begin{align}
    C_{a,b}\ket{n_a}_a\ket{n_b}_b
    =
    \ket{n_a}_a\ket{n_a\oplus n_b}_b\, . 
\end{align}
In our convention, the Pauli-$Z$ gate acts on a qubit as
\begin{align}
    Z_a\ket{n_a}=(-1)^{n_a}\ket{n_a}.
\end{align}
Therefore, 
\begin{align}
(C_{a,b}Z_bC_{a,b})\ket{n_a}_a\ket{n_b}_b
    &=
C_{a,b}Z_b\ket{n_a}_a\ket{n_a\oplus n_b}_b
\nonumber\\
    &=
C_{a,b}(-1)^{n_a\oplus n_b}\ket{n_a}_a\ket{n_a\oplus n_b}_b
\nonumber\\
    &=
(-1)^{n_a\oplus n_b}\ket{n_a}_a\ket{n_b}_b
\nonumber\\
    &=
(Z_a\otimes Z_b)\ket{n_a}_a\ket{n_b}_b\, ,  
\end{align}
implying
\begin{align}
Z_a\otimes Z_b=C_{a,b}Z_bC_{a,b}\, .     
\end{align}
Using this relation multiple times, one can show the following relations:
\begin{subequations}
\begin{align}
    &Z_{a_1}Z_{a_2}\cdots Z_{a_n} 
    =
    \left(\prod_{i=1}^{n-1}C_{a_i,a_n}\right)
    \cdot Z_{a_n}\cdot
    \left(\prod_{i=1}^{n-1}C_{a_i,a_n}\right),\\
    &\exp\left(-\mathrm{i}\theta\mathcal{C}'_{a_1\cdots a_n} Z_{a_1}Z_{a_2}\cdots Z_{a_n} \right)
    =
    \left(\prod_{i=1}^{n-1}C_{a_i,a_n}\right)
    \cdot \exp\left(-\mathrm{i}\theta\mathcal{C}'_{a_1\cdots a_n} Z_{a_n}\right)\cdot
   \left(\prod_{i=1}^{n-1}C_{a_i,a_n}\right).
\end{align}    
\end{subequations}

The decomposition into CNOT gates is not unique. We could also have, for example, 
\begin{align}
    Z_{a_1}Z_{a_2}\cdots Z_{a_n}
    =
    \left(C_{a_1,a_2}\cdots C_{a_{n-1},a_n}\right)
    \cdot Z_{a_n}\cdot
    \left(C_{a_{n-1},a_n}\cdots C_{a_1,a_2}\right)\, . 
\end{align}
Using these relations, we can write each term in the products \eqref{eq:interaction-generic-Z} and \eqref{eq: diagonal kinetic terms} in terms of CNOT gates and a one-qubit rotation. 

%%%%%%%%%%%%%%%%%%%%%%
%%%%%%%%%%%%%%%%%%%%%%
\subsubsection{Gate complexity}
%%%%%%%%%%%%%%%%%%%%%%
%%%%%%%%%%%%%%%%%%%%%%
Let us consider QFT with local interaction defined on a lattice, using noncompact variables. Let the lattice size be $L^d$. The number of terms in the kinetic and interaction parts is proportional to $L^d$. If the interaction term is a polynomial of degree $n$, there are $\sim L^dQ^n$ couplings of $n$ Pauli $Z$'s, and we will need $\sim nL^dQ^n$ CNOT gates and $\sim L^dQ^n$ one-qubit rotations. With a natural assumption $n>2$, the cost for the kinetic part and quantum Fourier transform is negligible in the limit of $Q\to\infty$. See Ref.~\cite{Halimeh:2024bth} for details. 

On noisy intermediate-scale quantum (NISQ) hardware, the number of CNOT gates is a reasonable indicator of the simulation cost. 
For fault-tolerant quantum computing (FTQC), it is important to reduce the number of T gates. For generic rotation angles, each one-qubit rotation costs roughly the same number of T gates, within the range of 10 -- 50~\cite{Campbell:2020wqh}. Therefore, whether we consider NISQ or FTQC, the cost scales proportionally to the volume $L^d$ and polynomially with respect to the number of qubits per boson $Q$. 

Note that we could count the gates and establish a polynomial scaling regarding $Q$ because the truncated Hamiltonian is so simple that we can write the circuit explicitly by hand. Such a robust counting is impossible in the Kogut--Susskind approach unless the Kogut--Susskind Hamiltonian is described as a special limit of the orbifold lattice. For example, although the potential term in the vanilla Kogut--Susskind with compact variables is sparse in the electric basis, an efficient implementation of the circuit is not known, and there is no known way to find an efficient implementation systematically avoiding exponentially large cost. Therefore, the orbifold lattice Hamiltonian, combined with the universal framework, achieves exponential speedup~\cite{Hanada:2025yzx,bergner2025exponentialspeedupquantumsimulation}. 
%%%%%%%%%%%%%%%%%%%%%%
%%%%%%%%%%%%%%%%%%%%%%
\section{Introducing fermions via Jordan--Wigner transform}\label{sec:universal_framework_boson+fermion}
%%%%%%%%%%%%%%%%%%%%%%
%%%%%%%%%%%%%%%%%%%%%%

%%%%%%%%%%%%%%%%%%%%%%%%%%%%%%%%%%%%%%%%%%%%
%%%%%%%%%%%%%%%%%%%%%%%%%%%%%%%%%%%%%%%%%%%%
\subsection{Jordan--Wigner transform}
%%%%%%%%%%%%%%%%%%%%%%%%%%%%%%%%%%%%%%%%%%%%
%%%%%%%%%%%%%%%%%%%%%%%%%%%%%%%%%%%%%%%%%%%%
We are interested in systems containing bosons and fermions. We assign each qubit either to bosonic or fermionic degrees of freedom. The Hilbert space is a tensor product of boson and fermion Hilbert spaces:
\begin{align}
    \mathcal{H}
    =
    \mathcal{H}^{\rm (bos)}
    \otimes
    \mathcal{H}^{\rm (fer)}\, . 
\end{align}
Bosonic operators acting on $\mathcal{H}^{\rm (bos)}$ are constructed as explained in Sec.~\ref{sec:universal_framework_boson}. In this section, we will define fermionic operators acting on $\mathcal{H}^{\rm (fer)}$. 

Let us consider the simplest way to implement fermions on quantum computers: the Jordan--Wigner transform~\cite{Jordan_Wigner}. Real fermions $\hat{\psi}_{a}$ ($a=1,2,\cdots$) that satisfy $\{\hat{\psi}_{a},\hat{\psi}_{b}\}=2\delta_{ab}$ can be written as 
\begin{align}
    \hat{\psi}_1
    &=
    X_1\, ,
    \nonumber\\
    \hat{\psi}_2
    &=
    Y_1\, ,
    \nonumber\\  
    \hat{\psi}_3
    &=
    Z_1\otimes X_2\, ,
    \nonumber\\  
    \hat{\psi}_4
    &=
    Z_1\otimes Y_2\, ,
    \nonumber\\
    &\cdots
    \nonumber\\
    \hat{\psi}_{2j-1}
    &=
    Z_1\otimes\cdots\otimes Z_{j-1}\otimes X_j\, ,
    \nonumber\\
    \hat{\psi}_{2j}
    &=
    Z_1\otimes\cdots\otimes  Z_{j-1}\otimes Y_j\, .
    \label{Jordan_Wigner}
\end{align}
Here, $X_i$, $Y_i$, and $Z_i$ are Pauli $X$, $Y$, and $Z$ operators acting on the $i$-th qubit in $\mathcal{H}^{\rm (fer)}$. 
We can introduce a complex fermion as
\begin{align}
    \hat{\Psi}_j = \hat{\psi}_{2j-1} + \mathrm{i}\hat{\psi}_{2j}\, , 
    \qquad
        \hat{\Psi}_j^\dagger = \hat{\psi}_{2j-1} - \mathrm{i}\hat{\psi}_{2j}\, .
\end{align}

We know that $\hat{x}$ is written as a weighted sum of Pauli $Z$'s in the coordinate basis. Therefore, terms in $V_{\rm fer}(\hat{x},\hat{\psi})$ can be written as a sum of  Pauli strings. The differences from the case of bosonic theories discussed in Sec.~\ref{sec:universal_framework_boson} are (i) $X$ and $Y$ appear both, and (ii) the Pauli strings can become long because fermions contain a long chain of Pauli $Z$'s. The second point is usually seen as problematic because it can lead to a larger gate count. In the treatment above, one $Z$ could be converted into two CNOTs, and hence many $Z$'s are converted to many CNOTs. As we will see, however, most CNOT gates cancel out in the case of local fermion bilinear interactions. 

To deal with (i), we can use 
\begin{align}
    X = hZh\, 
    \qquad
    Y = -s^\dagger hZhs\, 
    \label{X->Z,Y->Z}
\end{align}
where $h$ and $s$ are the Hadamard gate and phase gate, respectively:
\begin{align}
    h
    =
    \frac{1}{\sqrt{2}}
    \left(
        \begin{array}{cc}
            1 & 1\\
            1 & -1
        \end{array}
    \right)\, , 
    \qquad
    s
    =
    \left(
        \begin{array}{cc}
            1 & 0\\
            0 & \mathrm{i}
        \end{array}
    \right)\, .      
\end{align}
As an example, let us consider $X_1\otimes Y_2\otimes Z_3\otimes\cdots\otimes Z_n$.
In the case of Hamiltonian time evolution via Trotterization, we can use 
\begin{align}
&
\exp\left(-\mathrm{i}\epsilon
X_1\otimes Y_2\otimes Z_3\otimes\cdots\otimes Z_n
\right)
\nonumber\\
&\qquad=
h_1s_2^\dagger h_2
\exp\left(\mathrm{i}\epsilon
Z_1\otimes Z_2\otimes Z_3\otimes\cdots\otimes Z_n
\right)
h_2s_2h_1
\nonumber\\
&\qquad=
h_1s_2^\dagger h_2
\left(\prod_{j=1}^{n-1}C_{j,n}\right)
\exp\left(\mathrm{i}\epsilon
Z_n   
\right)
\left(\prod_{j=1}^{n-1}C_{j,n}\right)
h_2s_2h_1.
\end{align}
We can further rewrite it as
\begin{align}
&
\exp\left(-\mathrm{i}\epsilon
X_1\otimes Y_2\otimes Z_3\otimes\cdots\otimes Z_n
\right)
\nonumber\\
&\qquad=
\left(\prod_{j=3}^{n-1}C_{j,n}\right)
h_1s_2^\dagger h_2
\left(C_{1,n}C_{2,n}\right)
\exp\left(\mathrm{i}\epsilon
Z_n   
\right)
\left(C_{1,n}C_{2,n}\right)
h_2s_2h_1
\left(\prod_{j=3}^{n-1}C_{j,n}\right).
\end{align}

Associated with a long chain of $Z$'s, many CNOT gates appear. As we will see in Sec.~\ref{sec:simplification_JW}, however, most of the CNOT gates in $\prod_{j=3}^{n-1}C_{j,n}$ cancel in theories with local fermion bilinear interactions, including QCD. 
%%%%%%%%%%%%%%%%%%%%%%
%%%%%%%%%%%%%%%%%%%%%%
\subsection{Simplification in Trotterization in theories with local fermion bilinear interactions}\label{sec:simplification_JW}
%%%%%%%%%%%%%%%%%%%%%%
%%%%%%%%%%%%%%%%%%%%%%
Let us consider an $L^d$ square lattice. We assume that there is one complex (two real fermions) on each site.\footnote{The following argument can be generalized to any even number of fermions on each site. In the case of QCD in $3+1$D, each quark corresponds to four complex fermions at each site. See Sec.~\ref{sec:gate_counting_QCD_JW} for details.} We use $\vec{n}=(n_1,\cdots,n_d)$, where $n_i=1,\cdots,L$, to specify a lattice point. We introduce an ordering $\vec{n}\prec\vec{n}'$ among lattice points, in such a way that $\vec{n}\prec\vec{n}'$ means 
$n_d<n'_d$, or
$n_d=n'_d$, $n_{d-1}<n'_{d-1}$, or
$n_d=n'_d$, $\cdots$,  $n_{i+1}=n'_{i+1}$, $n_{i}<n'_{i}$, all the way down to $i=1$. We assign a number from 1 to $L^d$ to lattice sites following this ordering; 1 for $\vec{n}=(1,\cdots,1)$ and $L^d$ for $\vec{n}=(L,\cdots,L)$.  
Using the Jordan--Wigner transform, the $j$-th fermion in this ordering can be written using
$Z_1\otimes\cdots Z_{j-1}\otimes X_j$ and $Z_1\otimes\cdots Z_{j-1}\otimes Y_j$. 

We consider the nearest-neighbor fermion bilinear interaction. Furthermore, we assume the periodic boundary condition. 
Another ordering of lattice points used in Sec.~\ref{sec:Verstraete-Cirac} leads to a similar simplification; see Appendix~\ref{sec:JW_another_ordering}. 
%%%%%%%%%%%%%%%%%%%%%%
%%%%%%%%%%%%%%%%%%%%%%
\subsubsection{One spatial dimension}
%%%%%%%%%%%%%%%%%%%%%%
%%%%%%%%%%%%%%%%%%%%%%
We start with the case of one dimension. We impose the periodic boundary condition and label the lattice points by $n=1,2,\cdots,L$. 
We assume nearest-neighbor interactions of the form
\begin{align}    \hat{\Psi}_n^{(\dagger)}\hat{\Psi}_{n+1}^{(\dagger)}\hat{O}^{\rm (bos)}_n\, , 
\end{align}
where $\hat{O}^{\rm (bos)}_n$ is made of operators acting on $\mathcal{H}^{\rm (bos)}$. In our universal framework, this is a tensor product of Pauli $ Z$'s acting on $\mathcal{H}^{\rm (bos)}$. In the case of QCD formulated using orbifold lattice~\cite{Bergner:2024qjl}, this is simply a few Pauli $Z$'s. 
$\hat{\Psi}_n^{(\dagger)}$ means either the annihilation operator $\hat{\Psi}_n$ or creation operator $\hat{\Psi}_n^{\dagger}$ of a complex fermion at site $n$. 
Using the Jordan--Wigner transform, these interaction terms are written as a sum of terms of the form
\begin{align}
    \sigma_n\otimes\sigma_{n+1}\otimes\hat{O}^{\rm (bos)}_n
\end{align}
for $n=1,2,\cdots,L-1$ and 
\begin{align}
    \sigma_1
    \otimes
    Z_2\otimes\cdots\otimes Z_{L-1}
    \otimes
    \sigma_{L}\otimes\hat{O}^{\rm (bos)}_L
\end{align}
for $n=L$. Here, $\sigma$ means $X$ or $Y$. 
A nontrivial long Pauli string appears only for $L=n$. 
When we consider Hamiltonian time evolution via Trotter decomposition, we can convert it to $\mathcal{O}(L)$ CNOT gates, using e.g., 
\begin{align}
&
    \exp\left(-\mathrm{i}\epsilon\sigma_1
    \otimes
    Z_2\otimes\cdots\otimes Z_{L-1}
    \otimes
    \sigma_{L}\otimes\hat{O}^{\rm (bos)}_L\right)
    \nonumber\\
    &\qquad=
    \left(\prod_{j=2}^{L-2}C_{j,L-1}\right)
        \exp\left(-\mathrm{i}\epsilon\sigma_1
    \otimes
    Z_{L-1}
    \otimes
    \sigma_{L}\otimes\hat{O}^{\rm (bos)}_L\right)
    \left(\prod_{j=2}^{L-2}C_{j,L-1}\right)\, . 
\end{align}
We can rewrite this expression further. 
For example, suppose $\sigma_1=X_1$, $\sigma_L=X_L$, and $\hat{O}^{\rm (bos)}_L=Z^{\rm (bos)}_aZ^{\rm (bos)}_bZ^{\rm (bos)}_cZ^{\rm (bos)}_d$, where $Z^{\rm (bos)}_{a,b,c,d}$ are Pauli $Z$ operators acting on $\mathcal{H}^{\rm (bos)}$.  Then, 
\begin{align}
&
        \exp\left(-\mathrm{i}\epsilon X_1
    \otimes
    Z_{L-1}
    \otimes
    X_{L}\otimes\hat{O}^{\rm (bos)}_L\right)
    \nonumber\\
    &\qquad=
    h_1h_L
\exp\left(-\mathrm{i}\epsilon
    Z_1 \otimes Z_{L-1}\otimes Z_{L}\otimes
    Z^{\rm (bos)}_a\otimes Z^{\rm (bos)}_b\otimes Z^{\rm (bos)}_c \otimes Z^{\rm (bos)}_d
    \right)
    h_1h_L
    \nonumber\\
        &\qquad=
    h_1h_L
    C_{L-1,1}C_{L,1}C_{a,1}C_{b,1}C_{c,1}C_{d,1}
\exp\left(-\mathrm{i}\epsilon
    Z_1
    \right)
    C_{L-1,1}C_{L,1}C_{a,1}C_{b,1}C_{c,1}C_{d,1}
    h_1h_L\, , 
\end{align}
where $h_1$ and $h_L$ are Hadamard gates. 

The conjugation by $\prod_{j=2}^{L-2}C_{j,L-1}$ may not be ideal because the depth of the circuit is of order $L$. 
We can use the technique explained in Sec.~\ref{sec:rearrangement_of_product} to rearrange it into a different form and reduce the depth to $\mathcal{O}(\log L)$. 
%%%%%%%%%%%%%%%%%%%%%%
%%%%%%%%%%%%%%%%%%%%%%
\subsubsection{Two spatial dimensions}
\begin{figure}[t!]
	\centering
    \includegraphics[width=0.75\textwidth]{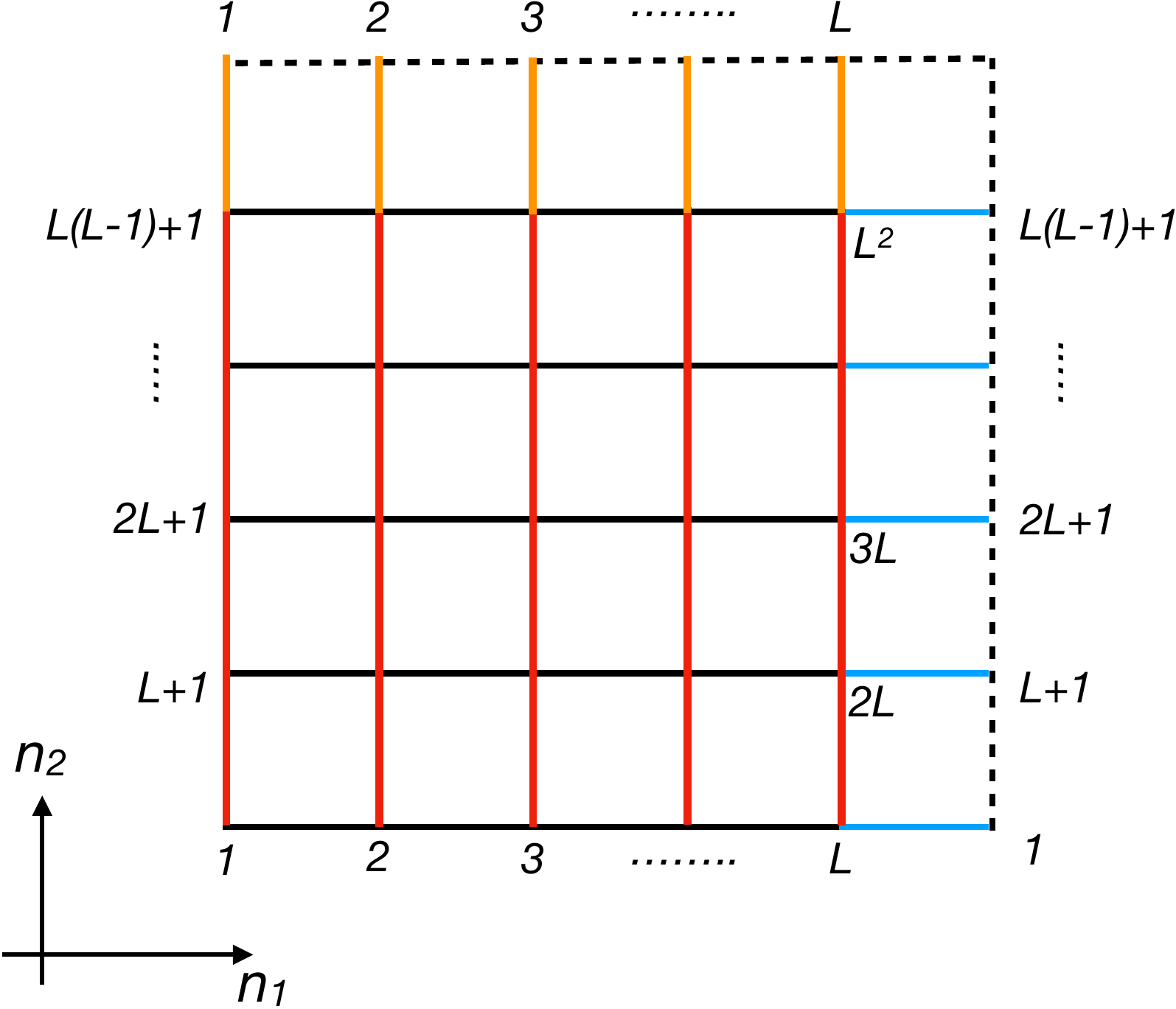}
	\caption{Two-dimensional $L\times L$ lattice with periodic boundary conditions. $L^2$ points are ordered following $\prec$.
    }
	\label{fig:2d-lattice}
\end{figure}

%%%%%%%%%%%%%%%%%%%%%%
%%%%%%%%%%%%%%%%%%%%%%
Next, we consider a two-dimensional lattice. We treat four types of links, shown in black, blue, red, and orange in Fig.~\ref{fig:2d-lattice}, separately. 

We already know how to deal with black and blue links, based on the argument for the 1D lattice. To see how we should treat red links, we consider $(1,L+1)$, $(2,L+2)$, ..., $(L,2L)$ together. 
Schematically, interaction terms are written as
\begin{align}
V_j\equiv
    \sigma_j
    \otimes
    Z_{j+1}\otimes\cdots\otimes Z_{j+L-1}
    \otimes
    \sigma_{j+L}
    \otimes
    \hat{O}^{\rm (bos)}_j\, , 
\end{align}
for $j=1,\cdots,L$. 
We consider the Trotter decomposition. 
For $j=2,\cdots,L$, we can convert a Pauli string to CNOT gates as
\begin{align}
\exp\left(
-\mathrm{i}\epsilon V_2
\right)
&=
C_{3,L+1}
C_{4,L+1}
\cdots
C_{L,L+1}
\nonumber\\
&\qquad\times
\exp\left(
-\mathrm{i}\epsilon 
    \sigma_2
    \otimes
    Z_{L+1}
    \otimes
    \sigma_{L+2}
    \otimes
    \hat{O}^{\rm (bos)}_j
\right)
\nonumber\\
&\qquad\times
C_{3,L+1}
\textcolor{red}{
C_{4,L+1}
\cdots
C_{L,L+1}}\, , 
\end{align}
%%%%
\begin{align}
\exp\left(
-\mathrm{i}\epsilon V_3
\right)
&=
\textcolor{red}{
C_{4,L+1}
\cdots
C_{L,L+1}}
\cdot
C_{L+2,L+1}
\nonumber\\
&\qquad\times
\exp\left(
-\mathrm{i}\epsilon 
    \sigma_3
    \otimes
    Z_{L+1}
    \otimes
    \sigma_{L+3}
    \otimes
    \hat{O}^{\rm (bos)}_j
\right)
\nonumber\\
&\qquad\times
C_{4,L+1}
\textcolor{red}{
C_{5,L+1}
\cdots
C_{L,L+1}
\cdot
C_{L+2,L+1}}\, , 
\end{align}
%%%%
\begin{align}
\exp\left(
-\mathrm{i}\epsilon V_j
\right)
&=
\textcolor{red}{C_{j+1,L+1}
\cdots
C_{L,L+1}
\cdot
C_{L+2,L+1}
\cdots
C_{L+j-2,L+1}
}
C_{L+j-1,L+1}
\nonumber\\
&\qquad\times
\exp\left(
-\mathrm{i}\epsilon 
    \sigma_j
    \otimes
    Z_{L+1}
    \otimes
    \sigma_{j+L}
    \otimes
    \hat{O}^{\rm (bos)}_j
\right)
\nonumber\\
&\qquad\times
C_{j+1,L+1}
\textcolor{red}{
C_{j+2,L+1}
\cdots
C_{L,L+1}
\cdot
C_{L+2,L+1}
\cdots
C_{L+j-1,L+1}}\, , 
\end{align}
%%%%
\begin{align}
\exp\left(
-\mathrm{i}\epsilon V_L
\right)
&=
\textcolor{red}{C_{L+2,L+1}
\cdots
C_{2L-2,L+1}
}
C_{2L-1,L+1}
\nonumber\\
&\qquad\times
\exp\left(
-\mathrm{i}\epsilon 
    \sigma_L
    \otimes
    Z_{L+1}
    \otimes
    \sigma_{2L}
    \otimes
    \hat{O}^{\rm (bos)}_L
\right)
\nonumber\\
&\qquad\times
C_{L+2,L+1}
\cdots
C_{2L-1,L+1}\, . 
\end{align}
If we take the product
$
\exp\left(
-\mathrm{i}\epsilon V_2
\right)
\cdot
\exp\left(
-\mathrm{i}\epsilon V_3
\right)
\cdot
\ldots
\cdot
\exp\left(
-\mathrm{i}\epsilon V_L
\right)
$, CNOT gates denoted in red cancel out, leaving only $\mathcal{O}(L)$ CNOT gates. Note that terms sandwiched by chains of CNOT gates can be expressed using $\mathcal{O}(1)$ CNOT gates, Hadamard gates, phase gates, and one-qubit rotations. In total, we need $\mathcal{O}(L)$ CNOT gates, Hadamard gates, phase gates, and one-qubit rotations for a product $
\exp\left(
-\mathrm{i}\epsilon V_2
\right)
\cdot
\exp\left(
-\mathrm{i}\epsilon V_3
\right)
\cdots
\exp\left(
-\mathrm{i}\epsilon V_L
\right)
$. 

We can have a similar cancellation of CNOT gates between the links $(1,L+1)$ and $(L,1)$. 

Other links written in red can be treated in the same way, leaving $\mathcal{O}(L^2)$ CNOT gates, Hadamard gates, phase gates, and one-qubit rotations in total. 

Finally, we consider orange links, i.e., $(1,L^2-L+1)$, $(2,L^2-L+2)$, ..., $(L,L^2)$. 
Schematically, interaction terms are written as 
\begin{align}
V'_j\equiv
    \sigma_j
    \otimes
    Z_{j+1}\otimes\cdots\otimes Z_{L^2-L+j-1}
    \otimes
    \sigma_{L^2-L+j}
    \otimes
    \hat{O}^{\prime\rm (bos)}_j\, , 
\end{align}
for $j=1,\cdots,L$. 
We can convert every Pauli string into CNOT gates as
\begin{align}
\exp\left(
-\mathrm{i}\epsilon V'_1
\right)
&=
C_{2,L+1}
\cdots
C_{L,L+1}
\cdot
C_{L+2,L+1}
\cdots
C_{L^2-L,L+1}
\nonumber\\
&\qquad\times
\exp\left(
-\mathrm{i}\epsilon 
    \sigma_1
    \otimes
    Z_{L+1}
    \otimes
    \sigma_{L^2-L+1}
    \otimes
    \hat{O}^{\prime\rm (bos)}_1
\right)
\nonumber\\
&\qquad\times
C_{2,L+1}
\textcolor{orange}{
C_{3,L+1}
\cdots
C_{L,L+1}
\cdot
C_{L+2,L+1}
\cdots
C_{L^2-L,L+1}}\, , 
\end{align}
%%%%
\begin{align}
\exp\left(
-\mathrm{i}\epsilon V'_2
\right)
&=
\textcolor{orange}{
C_{3,L+1}
\cdots
C_{L,L+1}
\cdot
C_{L+2,L+1}
\cdots
C_{L^2-L,L+1}
}
C_{L^2-L+1,L+1}
\nonumber\\
&\qquad\times
\exp\left(
-\mathrm{i}\epsilon 
    \sigma_2
    \otimes
    Z_{L+1}
    \otimes
    \sigma_{L^2-L+2}
    \otimes
    \hat{O}^{\prime\rm (bos)}_2
\right)
\nonumber\\
&\qquad\times
C_{3,L+1}
\textcolor{orange}{
C_{4,L+1}
\cdots
C_{L,L+1}
\cdot
C_{L+2,L+1}
\cdots
C_{L^2-L+1,L+1}}\, , 
\end{align}
%%%%
\begin{align}
\exp\left(
-\mathrm{i}\epsilon V'_j
\right)
&=
\textcolor{orange}{C_{j+1,L+1}
\cdots
C_{L,L+1}
\cdot
C_{L+2,L+1}
\cdots
C_{L^2-L+j-2,L+1}
}
C_{L^2-L+j-1,L+1}
\nonumber\\
&\qquad\times
\exp\left(
-\mathrm{i}\epsilon 
    \sigma_j
    \otimes
    Z_{L+1}
    \otimes
    \sigma_{L^2-L+j}
    \otimes
    \hat{O}^{\prime\rm (bos)}_2
\right)
\nonumber\\
&\qquad\times
C_{j+1,L+1}
\textcolor{orange}{
C_{j+2,L+1}
\cdots
C_{L,L+1}
\cdot
C_{L+2,L+1}
\cdots
C_{L^2-L+j-1,L+1}}\, , 
\end{align}
%%%%
\begin{align}
\exp\left(
-\mathrm{i}\epsilon V'_L
\right)
&=
\textcolor{orange}{
C_{L+2,L+1}
\cdots
C_{L^2-2,L+1}}
C_{L^2-1,L+1}
\nonumber\\
&\qquad\times
\exp\left(
-\mathrm{i}\epsilon 
    \sigma_L
    \otimes
    Z_{L+1}
    \otimes
    \sigma_{L^2}
    \otimes
    \hat{O}^{\prime\rm (bos)}_L
\right)
\nonumber\\
&\qquad\times
C_{L+2,L+1}
\cdots
C_{L^2-1,L+1}\, . 
\end{align}
If we take a product
$
\exp\left(
-\mathrm{i}\epsilon V'_1
\right)
\cdot
\exp\left(
-\mathrm{i}\epsilon V'_2
\right)
\cdots
\exp\left(
-\mathrm{i}\epsilon V'_L
\right)
$, CNOT gates denoted by orange cancel out, leaving only $\mathcal{O}(L^2)$ CNOT gates and $\mathcal{O}(L)$ Hadamard gates, phase gates, and one-qubit rotations.

%%%%%%%%%%%%%%%%%%%%%%
%%%%%%%%%%%%%%%%%%%%%%
\subsubsection{$d>2$ spatial dimensions}
%%%%%%%%%%%%%%%%%%%%%%
%%%%%%%%%%%%%%%%%%%%%%
The generalization to higher dimensions is straightforward. Associated with links along the $d$-th dimension, Pauli strings with length $L^{d-1}$ appear. They can be converted to CNOT gates, which cancel out just as in the case of $d=2$. To see it systematically, we can think as though the first $d-1$ dimensions as the `first dimension' consisting of $L^{d-1}$ points, and the $d$-th dimension as the `second dimension' consisting of $L$ points. 

After the cancellation, only $\mathcal{O}(L^d)$ CNOT gates remain. We also have $\mathcal{O}(L^d)$ numbers of one-qubit gates, including one-qubit rotations. 

If we do not utilize the cancellation pattern shown above, there will be $\mathcal{O}(L^{2d-1})$ CNOT gates. 

%%%%%%%%%%%%%%%%%%%%%%
%%%%%%%%%%%%%%%%%%%%%%
\subsubsection{Rearranging $\prod_{j=1}^n C_{j,n+1}$}\label{sec:rearrangement_of_product}
%%%%%%%%%%%%%%%%%%%%%%
%%%%%%%%%%%%%%%%%%%%%%
In the above, most of the long chains of CNOT gates were canceled, but a few long chains of the form $\prod_{j=1}^K C_{j,K+1}$ remained. The same sort of chain appears associated with an operator insertion that is not involved with the cancellation of CNOT gates. 
These products are somewhat problematic in the sense that the operations of the gates cannot be parallelized (equivalently, the depth is $K$), because all CNOT gates share the same target qubit. 

\begin{figure}[t!]
	\centering
\includegraphics[width=10cm]{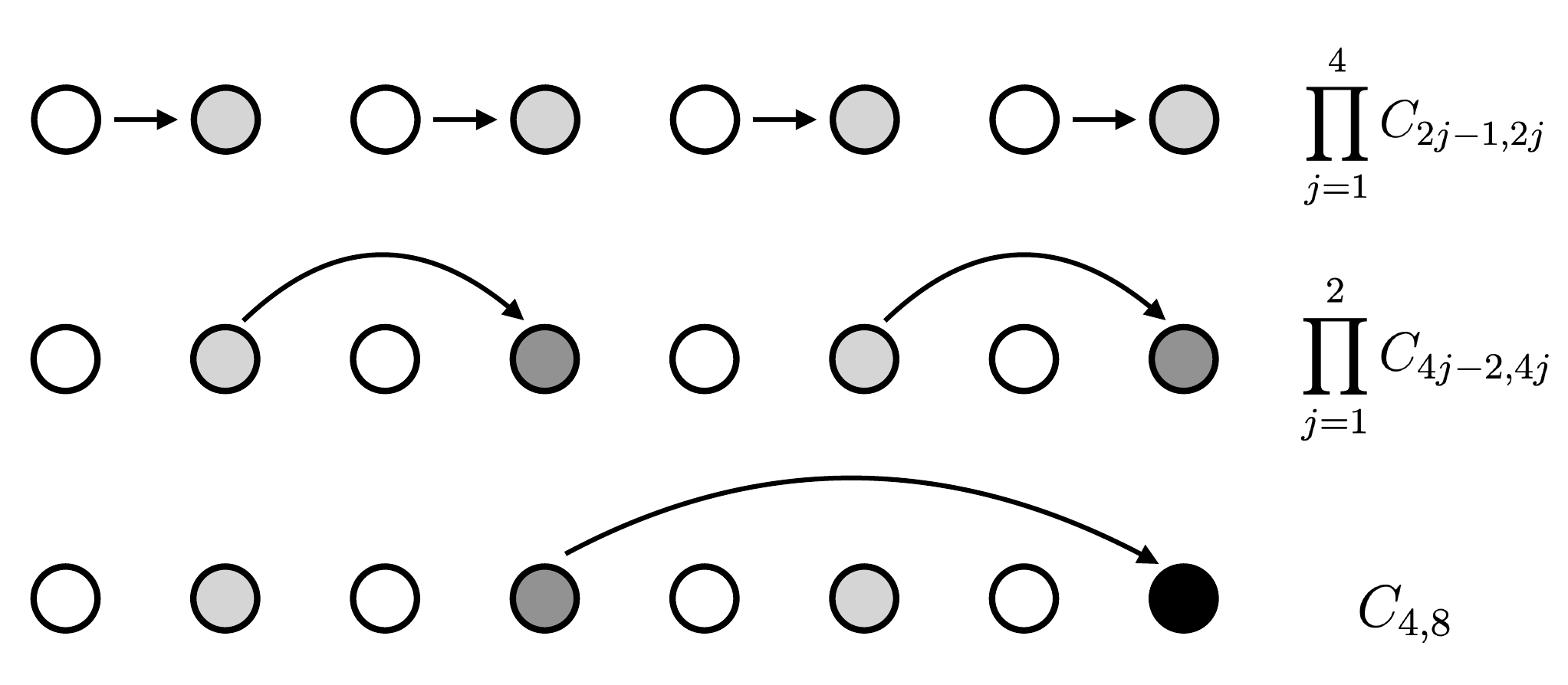}
	\caption{
    This figure explains the final line in \eqref{eq:CNOT_rearrange} visually for $\ell=3$. Each arrow indicates the action of a CNOT gate. 
    Acting these operations to $\prod_{a=1}^8\ket{n_a}_a$, we turn the 8-th qubit to be in the state $\ket{n_1\oplus\cdots\oplus n_8}_8$. 
    While $2^\ell-1=7$ CNOT gates are needed, the depth is $\ell=3$. 
    }
\label{fig:CNOT_rearrange}
\end{figure}

This problem can be resolved by rearranging the products in the following manner. For simplicity, let $K=2^\ell+1$. 
Then, using 
\begin{align}
\left(
    \prod_{j=1}^{2^{\ell-1}}C_{2j-1,2j}
    \right)
    \left(
    \otimes_{j=1}^{2^\ell}
 \ket{n_j}_j
 \right)
 =
    \otimes_{j=1}^{2^{\ell-1}}
    \left(
    \ket{n_{2j-1}}_{2j-1}
    \ket{n_{2j-1}\oplus n_{2j}}_{2j} 
    \right)\, , 
\end{align}
\begin{align}
&
\left(
\prod_{j=1}^{2^{\ell-2}}C_{4j-2,4j}
\right)
    \otimes_{j=1}^{2^{\ell-1}}
    \left(
    \ket{n_{2j-1}}_{2j-1}
    \ket{n_{2j-1}\oplus n_{2j}}_{2j} 
    \right)
\nonumber\\
&\qquad=
    \otimes_{j=1}^{2^{\ell-2}}
    \left(
    \ket{n_{4j-3}}_{4j-3}
    \ket{n_{4j-3}\oplus n_{4j-2}}_{4j-2} 
    \ket{n_{4j-1}}_{4j-1}
    \ket{n_{4j-3}\oplus n_{4j-2}\oplus n_{4j-1}\oplus n_{4j}}_{4j} 
    \right)\, , 
\end{align}
etc, we can see that
\begin{align}
    \prod_{j=1}^{2^\ell} C_{j,2^\ell+1}
    &=
    \left(
    \prod_{j=1}^{2^{\ell-1}}C_{2j-1,2j}
    \right)   
    \left(
    \prod_{j=1}^{2^{\ell-2}}C_{4j-2,4j}
    \right)
    \cdots
    \left(
    \prod_{j=1}^{2}C_{2^{\ell-1}j-2^{\ell-2},2^{\ell-1}j}
    \right)  
    C_{2^{\ell-1},2^\ell}
    \nonumber\\
    &
    \times C_{2^\ell,2^\ell+1}
        \nonumber\\
    &
    \times
    C_{2^{\ell-1},2^\ell}
    \left(
    \prod_{j=1}^{2}C_{2^{\ell-1}j-2^{\ell-2},2^{\ell-1}j}
    \right)  
    \cdots
    \left(
    \prod_{j=1}^{2^{\ell-2}}C_{4j-2,4j}
    \right)
        \left(
    \prod_{j=1}^{2^{\ell-1}}C_{2j-1,2j}
    \right)\, .    
    \label{eq:CNOT_rearrange}
\end{align}
See Fig.~\ref{fig:CNOT_rearrange}. Now the depth is $\mathcal{O}(\ell)=\mathcal{O}(\log K)$. For our purposes, $K\sim L^d$ and $\log K\sim d\log L$. 

This trick can be used for the block encoding in terms of a linear combination of unitaries; see Appendix~\ref{sec:block_encoding_JW}.

%%%%%%%%%%%%%%%%%%%%%%
%%%%%%%%%%%%%%%%%%%%%%
\subsubsection{QCD via Jordan--Wigner transform}\label{sec:gate_counting_QCD_JW}
%%%%%%%%%%%%%%%%%%%%%%
%%%%%%%%%%%%%%%%%%%%%%
We consider SU($N_{\rm c}$) QCD with $N_{\rm f}$ flavors of Dirac fermions in the fundamental representation on a three-dimensional spatial lattice. On each lattice point, we need to put multiple fermionic degrees of freedom. Each of $N_{\rm f}$ Dirac fermions has $N_{\rm c}$ color degrees of freedom, and each of them consists of eight real degrees of freedom. Therefore, $8N_{\rm c}N_{\rm f}$ real degrees of freedom are on each lattice site. In the following, we assume $N_{\rm c}$ and $N_{\rm f}$ are fixed\footnote{Typically, $N_{\rm c}=3$ and $N_{\rm f}=2$ (up quark and down quark) or $N_{\rm f}=3$ (up, down, and strange).} and focus on the scaling of the cost with lattice size $L$. 

We use the Jordan--Wigner transform \eqref{Jordan_Wigner} and the notation 
\begin{align}
\Psi_{n,a}\equiv\psi_{8N_{\rm c}N_{\rm f}(n-1)+a}\, , 
\label{quarks}
\end{align}
where $n=1,2,\cdots,L^3$ labels lattice site and $a=1,2,\cdots,8N_{\rm c}N_{\rm f}$ labels degrees of freedom at each site. When we use Trotterization for Hamiltonian time evolution, we can use the cancellation of CNOT gates discussed above if we order the interactions based on the ordering of the site label, $n$. Therefore, the quantum circuit can be written using $\mathcal{O}(L^3)$ CNOT gates, $\mathcal{O}(L^3)$ Hadamard and phase gates, and $\mathcal{O}(L^3)$ one-qubit rotations for each Trotter step. The number of T-gates is proportional to the number of one-qubit rotations. 

It is straightforward to introduce arbitrary matter content. The only difference is that the number of fermions at each site changes from $8N_{\rm c}N_{\rm f}$ to an appropriate number, depending on the representations of the matter fields. 
Generalization to other dimensions is also straightforward. The gate counting is $\mathcal{O}(L^d)$ CNOT gates, $\mathcal{O}(L^d)$ Hadamard and phase gates, and $\mathcal{O}(L^d)$ one-qubit rotations for each Trotter step. If we did not use the cancellations between the CNOT gates, we could have $\mathcal{O}(L^{2d-1})$ CNOT gates.
Interestingly, even though we are using the Jordan--Wigner transform, the gate counting is proportional to the volume, $L^d$. The nonlocality of the Jordan--Wigner transform does not cause an increase in the number of gates because of the cancellation of the CNOT gates. 
The remnant of the nonlocality is the specific ordering of the product needed for the cancellation of the CNOT gates. As we will see below, when the Verstraete--Cirac transform is used, the nonlocality is eliminated, and specific ordering is not needed. 
%%%%%%%%%%%%%%%%%%%%%%
%%%%%%%%%%%%%%%%%%%%%%
\section{Introducing fermions via Verstraete--Cirac transform
}\label{sec:Verstraete-Cirac}
%%%%%%%%%%%%%%%%%%%%%%
%%%%%%%%%%%%%%%%%%%%%%
The Jordan--Wigner transformation, while being the most commonly used fermion-to-qubit transformation in general dimensions, is fundamentally a one-dimensional mapping. In this approach, spins and fermions are ordered in a one-dimensional sequence, and the transformation preserves the anticommutation relations within this linear structure. In two or higher dimensions, local fermion operators are mapped into highly nonlocal qubit operators. Although we observed a significant simplification of the quantum circuit for the case of the Trotter time evolution, we do not find a similar simplification for the case of block encoding the Hamiltonian as a linear combination of unitaries, as we discuss in Appendix~\ref{sec:block_encoding_JW}. This motivates us to consider more sophisticated approaches than the Jordan--Wigner transform. 

To resolve the issues associated with the Jordan--Wigner transform, various proposals suitable for higher dimensions have been developed \cite{Bravyi2002AnPhy,Yu-An2018AnPhy,Verstraete-Cirac2005,Kitaev2006AnPhy,Whitfield_2016,Jiang2019PhRvP,Setia2019PhRvR,Bochniak2020JHEP,Derby_2021,Po2021arXiv210710842P,Chen:2018nog,Chen2020PhRvR,Chen2023PRXQ}. In this section, we consider the Verstraete--Cirac encoding scheme~\cite{Verstraete:2005pn}, which uses Jordan--Wigner transformation in its construction, but the resultant qubit operators are still local in two and higher dimensions. 

%%%%%%%%%%%%%%%%%%%%%%%%%%%%%%%%%%%%%%%%%%%%
%%%%%%%%%%%%%%%%%%%%%%%%%%%%%%%%%%%%%%%%%%%%
\subsection{Verstraete--Cirac transform}
%%%%%%%%%%%%%%%%%%%%%%%%%%%%%%%%%%%%%%%%%%%%
%%%%%%%%%%%%%%%%%%%%%%%%%%%%%%%%%%%%%%%%%%%%
The Verstraete--Cirac transform~\cite{Verstraete:2005pn} assumes a local structure with nearest-neighbor interaction. 
Let us first consider a square lattice in two spatial dimensions, and then generalize it to three spatial dimensions. Further generalization to higher dimensions is straightforward. We will use the open boundary condition instead of the periodic boundary condition. 
%%%%%%%%%%%%%%%%%%%%%%%%%%%%%%%%%%%%%%%%%%%%
%%%%%%%%%%%%%%%%%%%%%%%%%%%%%%%%%%%%%%%%%%%%
\subsubsection{Verstraete--Cirac transform in two spatial dimensions}\label{sec:VC-2d}
\begin{figure}[t!]
	\centering
    \includegraphics[width=10cm]{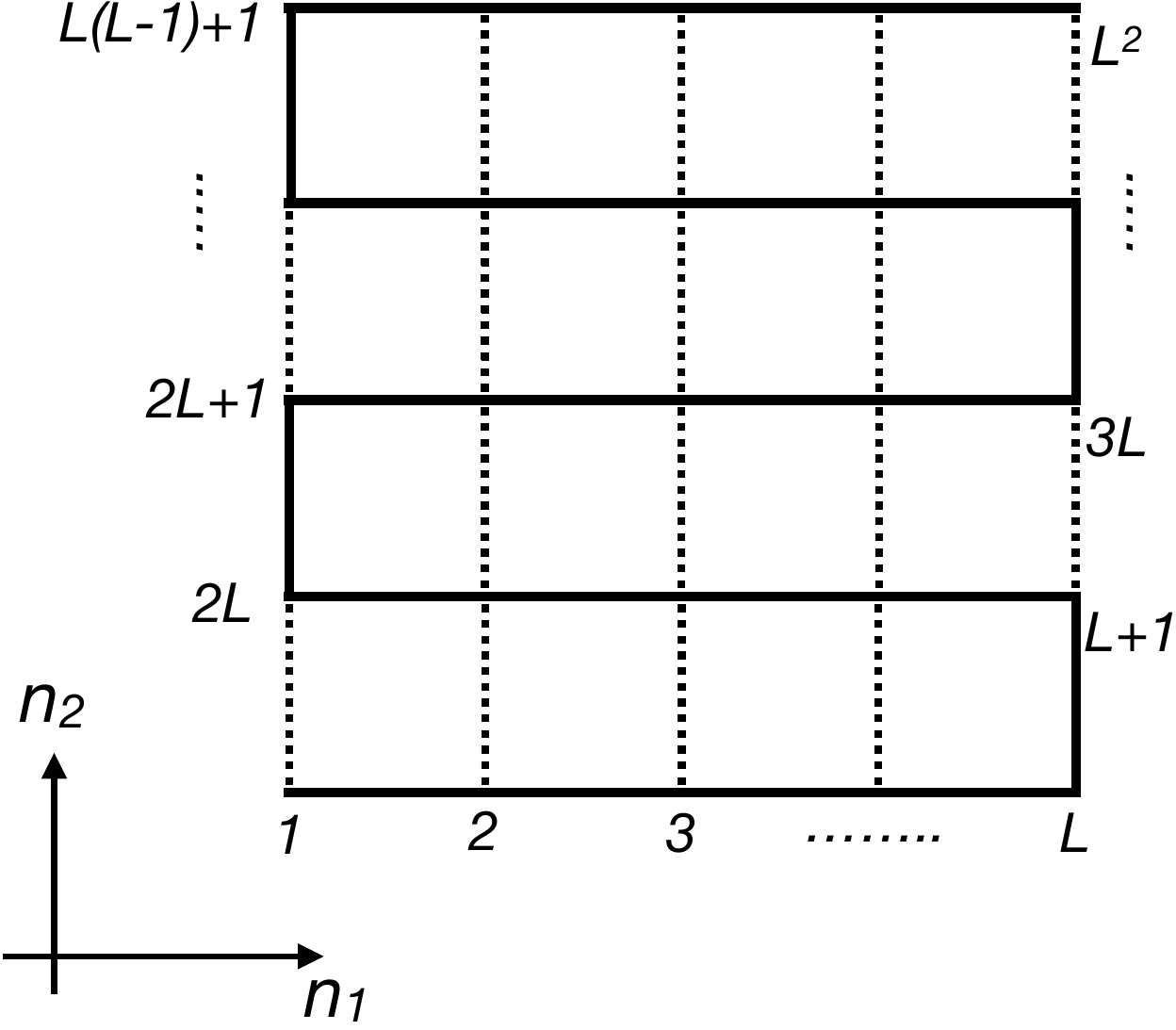}
	\caption{
    The ordering of lattice points for the Verstraete--Cirac transform in two spatial dimensions.
    }
	\label{fig:VC-2d}
\end{figure}

%%%%%%%%%%%%%%%%%%%%%%%%%%%%%%%%%%%%%%%%%%%%
%%%%%%%%%%%%%%%%%%%%%%%%%%%%%%%%%%%%%%%%%%%%
For the Verstraete--Cirac transform, we order the lattice points as shown in Fig.~\ref{fig:VC-2d}. 
To take into account `internal' degrees of freedom at each point (color, flavor, and spinor indices), we use the notation \eqref{quarks}. We assume nearest-neighbor interactions of the form
\begin{align}
\sum_{a,b}\hat{O}_{a,b}\hat{\Psi}_{n,a}\hat{\Psi}_{n',b}\, . 
\label{VC_original_interaction}
\end{align}
Here, $\hat{O}_{a,b}$ depends only on bosons and not on fermions. $n$ and $n'$ are neighboring points, e.g., 1 and 2, 1 and $2L$, or $2L$ and $2L+1$.  

Suppose that we used the standard Jordan--Wigner transform. Then, long chains of $Z$ gates appear when $n$ and $n'$ are separated to the second direction (i.e., when $n$ corresponds to $(n_1,n_2)$ and $n'$ corresponds to $(n_1,n_2+1)$). 
The key idea of the Verstraete--Cirac transform is to introduce auxiliary fermions that cancel those chains. Specifically, we can introduce real fermions $\hat{\rho}_{n}$ and $\hat{\chi}_{n}$, and replace \eqref{VC_original_interaction} with 
\begin{align}
\sum_{a,b}\hat{O}_{n,a;n',b}\hat{\Psi}_{n,a}\hat{\Psi}_{n',b}\cdot\mathrm{i}\hat{\rho}_{n}\hat{\chi}_{n'}
=
\mathrm{i}\sum_{a,b}\hat{O}_{n,a;n',b}
\left(\hat{\Psi}_{n,a}\hat{\rho}_{n}\right)
\left(\hat{\Psi}_{n',b}\hat{\chi}_{n'}\right)\, , 
\label{VC_improved_interaction}
\end{align}
\textit{when $n$ and $n'$ ($n<n'$) are neighboring points separated along the second direction, e.g.,} $(n,n')=(1,2L)$, $(2,2L-1)$, $(2L, 2L+1)$, or $(L, L+1)$. 
In the Jordan--Wigner transform, $\hat{\rho}_n$ and $\hat{\chi}_n$ are taken right after $\hat{\Psi}_{n,a}$. Then, neither $\hat{\Psi}_{n,a}\hat{\rho}_{n}$ nor $\hat{\Psi}_{n',b}\hat{\chi}_{n'}$ involve a long chain of $Z$s. 
Note that $\mathrm{i}\hat{\rho}_{n}\hat{\chi}_{n'}$ is related to the number operator of a complex fermion $\hat{c}_{n,n'}\equiv\frac{\hat{\rho}_{n}+\mathrm{i}\hat{\chi}_{n'}}{\sqrt{2}}$. Specifically, 
\begin{align}
    \hat{c}^\dagger_{n,n'}\hat{c}_{n,n'}
    =
    1-\mathrm{i}\hat{\rho}_{n}\hat{\chi}_{n'}\, . 
\end{align}
Therefore, if we restrict the states to be the Fock vacuum of these complex fermions, $\mathrm{i}\hat{\rho}_{n}\hat{\chi}_{n'}$ in \eqref{VC_improved_interaction} can be replaced with 1, and \eqref{VC_improved_interaction} reduces to \eqref{VC_original_interaction}. If $\mathrm{i}\hat{\rho}_{n}\hat{\chi}_{n'}$ is set to 1 as the initial condition, the Hamiltonian time evolution does not alter this condition. 

At this stage, the problem is still somewhat complicated because long Pauli strings can be associated with  $\mathrm{i}\hat{\rho}_{n}\hat{\chi}_{n'}$. To see how this issue can be resolved, let us discuss the case of $1\le n\le L$ as an example. We want to set 
$\mathrm{i}\hat{\rho}_{1}\hat{\chi}_{2L}$, 
$\mathrm{i}\hat{\rho}_{2}\hat{\chi}_{2L-1}$, 
$\mathrm{i}\hat{\rho}_{3}\hat{\chi}_{2L-2}$, 
..., and $\mathrm{i}\hat{\rho}_{L}\hat{\chi}_{L+1}$ to 1. This is equivalent to setting 
$\mathrm{i}\hat{\rho}_{1}\hat{\chi}_{2L}\cdot\mathrm{i}\hat{\rho}_{2}\hat{\chi}_{2L-1}$, 
$\mathrm{i}\hat{\rho}_{2}\hat{\chi}_{2L-1}\cdot\mathrm{i}\hat{\rho}_{3}\hat{\chi}_{2L-2}$, ..., 
$\mathrm{i}\hat{\rho}_{L-1}\hat{\chi}_{L+2}\cdot\mathrm{i}\hat{\rho}_{L}\hat{\chi}_{L+1}$, 
and $\mathrm{i}\hat{\rho}_{L}\hat{\chi}_{L+1}$ to be 1. 
Now, note that $\mathrm{i}\hat{\rho}_{L}\hat{\chi}_{L+1}$ is not associated with a long Pauli string, and 
$\mathrm{i}\hat{\rho}_{n}\hat{\chi}_{n'}\cdot\mathrm{i}\hat{\rho}_{n+1}\hat{\chi}_{n'-1}$ ($n=1,2,\cdots,L-1$) is also free from long Pauli strings, because
\begin{align}
\mathrm{i}\hat{\rho}_{n}\hat{\chi}_{n'}\cdot\mathrm{i}\hat{\rho}_{n+1}\hat{\chi}_{n'-1}
=
(\hat{\rho}_{n}\hat{\rho}_{n+1})
\cdot
(\hat{\chi}_{n'-1}\hat{\chi}_{n'})\, . 
\end{align}
Therefore, to set all $\mathrm{i}\hat{\rho}_{n}\hat{\chi}_{n'}$ in \eqref{VC_improved_interaction} to be 1 simultaneously, we can minimize 
\begin{align}
-
\left(
\mathrm{i}\hat{\rho}_{1}\hat{\chi}_{2L}\cdot\mathrm{i}\hat{\rho}_{2}\hat{\chi}_{2L-1}
+
\mathrm{i}\hat{\rho}_{2}\hat{\chi}_{2L-1}\cdot\mathrm{i}\hat{\rho}_{3}\hat{\chi}_{2L-2}
+
\cdots
+
\mathrm{i}\hat{\rho}_{L}\hat{\chi}_{L+1}
+
\cdots
\right)\, ,  
\end{align}
which is not associated with long Pauli strings. 
%%%%%%%%%%%%%%%%%%%%%%%%%%%%%%%%%%%%%%%%%%%%
%%%%%%%%%%%%%%%%%%%%%%%%%%%%%%%%%%%%%%%%%%%%
\subsubsection{Verstraete--Cirac transform in three spatial dimensions}\label{sec:VC-3d}
\begin{figure}[t!]
	\centering
    \includegraphics[width=10cm]{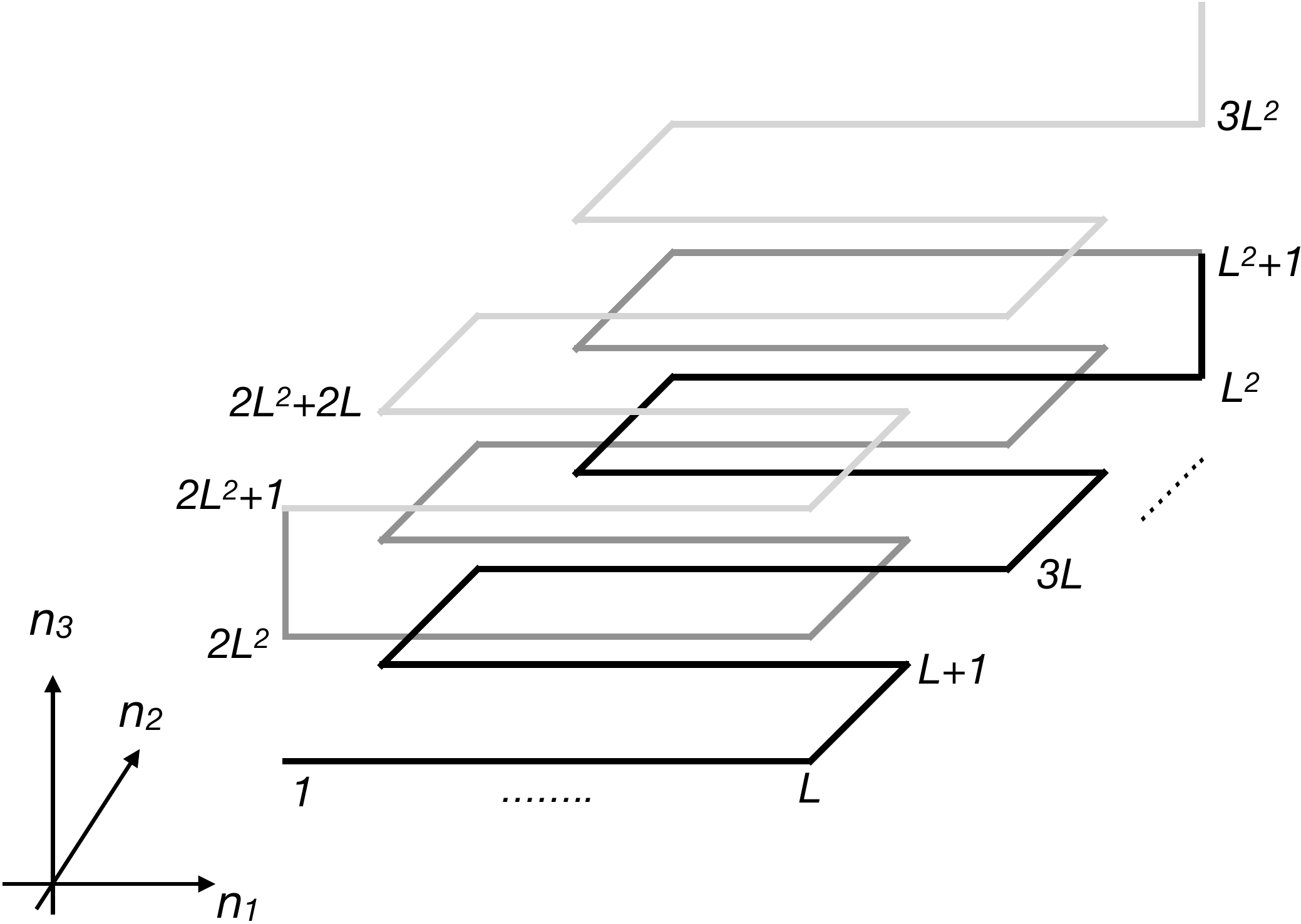}
	\caption{The ordering of lattice points for the Verstraete--Cirac transform in three spatial dimensions.
    }
	\label{fig:VC-3d}
\end{figure}

%%%%%%%%%%%%%%%%%%%%%%%%%%%%%%%%%%%%%%%%%%%%
%%%%%%%%%%%%%%%%%%%%%%%%%%%%%%%%%%%%%%%%%%%%
To generalize the Verstraete--Cirac transform to three spatial dimensions, we order the lattice points as shown in Fig.~\ref{fig:VC-3d}. To tame the coupling along the third direction, we regard the two-dimensional slice at fixed $n_3$ as though it is `one dimension' consisting of $L^2$ points, and repeat the same argument as the case of two dimensions. Specifically:
\begin{itemize}
\item 
For each fixed-$n_3$ two-dimensional slice, we eliminate long Pauli strings arising from the couplings along the second direction by adding auxiliary fermions $\rho^{(n_3)}_n$ and $\chi^{(n_3)}_n$, where $n=1,2,\cdots,L^2$, and repeat the same procedures as Sec.~\ref{sec:VC-2d}.  
\item 
To eliminate long Pauli strings arising from the couplings along the third direction, we introduce auxiliary fermions $\rho'_n$ and $\chi'_n$, where $n=1,2,\cdots,L^3$. The original coupling \eqref{VC_original_interaction} is replaced with 
\begin{align}
\sum_{a,b}\hat{O}_{n,a;n',b}\hat{\Psi}_{n,a}\hat{\Psi}_{n',b}\cdot\mathrm{i}\hat{\rho}'_{n}\hat{\chi}'_{n'} 
\end{align}
when $n$ and $n'$ ($n<n'$) are neighboring points separated along the third direction. We chose the states to be in the Fock vacuum of the auxiliary complex fermions $\hat{c}'_{n,n'}\equiv\frac{\hat{\rho}'_{n}+\mathrm{i}\hat{\chi}'_{n'}}{\sqrt{2}}$. 
\end{itemize}

%%%%%%%%%%%%%%%%%%%%%%%%%%%%%%%%%%%%%%%%%%%%%%
%%%%%%%%%%%%%%%%%%%%%%%%%%%%%%%%%%%%%%%%%%%%%%
\subsection{Universal framework with Verstraete--Cirac transform}\label{sec:universal_framework_VC}
%%%%%%%%%%%%%%%%%%%%%%%%%%%%%%%%%%%%%%%%%%%%%%
%%%%%%%%%%%%%%%%%%%%%%%%%%%%%%%%%%%%%%%%%%%%%%
If we describe fermions using Verstraete--Cirac transform while bosons are described as explained in Sec.~\ref{sec:universal_framework_boson}, then the kinetic part $\frac{1}{2}\sum_a\hat{p}_a^2$ does not change, while the potential term changes slightly: Pauli strings become longer (but stay $\mathcal{O}(L^0)$ in number, unlike the case of the Jordan--Wigner transform), and they contain $X$ and $Y$ gates in addition to $Z$.
%%%%%%%%%%%%%%%%%%%%%%
%%%%%%%%%%%%%%%%%%%%%%
\subsubsection{Trotter decomposition via Verstraete--Cirac transform}
%%%%%%%%%%%%%%%%%%%%%%
%%%%%%%%%%%%%%%%%%%%%%
Because the lengths of Pauli strings are of order $L^0$, Hamiltonian time evolution via Trotter decomposition is straightforward. Specifically, unlike the case of the Jordan--Wigner transform discussed Sec.~\ref{sec:simplification_JW}, we do not have to cancel long chains of CNOT gates; such chains are absent by construction.  
%%%%%%%%%%%%%%%%%%%%%%
%%%%%%%%%%%%%%%%%%%%%%
\subsubsection{Block encoding via Verstraete--Cirac transform}
%%%%%%%%%%%%%%%%%%%%%%
%%%%%%%%%%%%%%%%%%%%%%
The block encoding in terms of a linear combination of unitaries is not very efficient with Jordan--Wigner, as explained in Appendix~\ref{sec:block_encoding_JW}. On the contrary, an efficient realization of the block encoding is straightforward with the Verstraete--Cirac transform. Below, we will review the block encoding for bosonic theories discussed in Ref.~\cite{Hanada:2025yzx} and modify it slightly to include fermions.  

In the coordinate basis, the bosonic Hamiltonian \eqref{eq:Hamiltonian-boson} is written using Pauli strings $\hat{\Pi}_i$, $\hat{\tilde{\Pi}}_{j}$ and the quantum Fourier transforms $\hat{F}$ as 
\begin{align}
\hat{H}
=
\sum_{i} \alpha_{i} \hat{\Pi}_{i}
+
\sum_{j} \tilde{\alpha}_{j}
\hat{F}^\dagger \hat{\tilde{\Pi}}_{j}\hat{F}
, \qquad \alpha_{i}>0,\quad \tilde{\alpha}_{j} > 0\, .  
\label{Hamiltonian-as-sum-of-Pauli-chains}
\end{align}
The first and second terms on the right-hand side are the potential and kinetic terms, respectively.
Both $\hat{\Pi}_i$ and $\hat{\tilde{\Pi}}_{j}$ are short and made only from Pauli $Z$ gates. 

For the block encoding of the Hamiltonian into a unitary matrix, ancilla states $\ket{i}$, $\ket{\tilde{j}}$ are introduced, and a state $|G\rangle$ is defined by  
\begin{align}
\ket{G}  
=  
\sum_i g_i \ket{i}
+
\sum_i \tilde{g}_i \ket{\tilde{j}}\, , 
\end{align}
where
\begin{align}
|g_i|^2
=
\frac{\alpha_i}{\lambda}\, , 
\qquad
|\tilde{g}_j|^2
=
\frac{\tilde{\alpha}_j}{\lambda}\, , 
\qquad
\lambda
=
\sum_i\alpha_i
+
\sum_j\tilde{\alpha}_j\, . 
\end{align} 
Then, we prepare a unitary operator $\hat{U}$, acting on the Hilbert space extended with ancilla qubits, that satisfies 
\begin{align}
\frac{\hat{H}}{\lambda} = \Big( \bra{G} \otimes \hat{I} \Big)  \hat{U}  \Big( \ket{G} \otimes\hat{I} \Big)\, .
\label{eq:H_G_U}
\end{align}
Here, $\hat{I}$ is the identity operator acting on the original Hilbert space.
More explicitly: 
\begin{align}
\hat{U} 
=
\sum_i
\left(
\ket{i}\bra{i}
\otimes
\hat{\Pi}_i
\right)
+
\hat{F}^\dagger
\sum_j
\left(
\ket{\tilde{j}}\bra{\tilde{j}}
\otimes\hat{\tilde{\Pi}}_j
\right)
\hat{F}\, . 
\end{align}
This can be written as
\begin{align}
\hat{U} 
=
\hat{F}^\dagger
\hat{U}_{\rm kin}
\hat{F}
\hat{U}_{\rm pot}\, ,  
\end{align}
where
\begin{align}
\hat{U}_{\rm pot}
=
\sum_i
\left(
\ket{i}\bra{i}
\otimes
\hat{\Pi}_i
\right)
+
\left(
\sum_j
\ket{\tilde{j}}\bra{\tilde{j}}
\otimes\hat{I}
\right)\, , 
\end{align}
%%%
\begin{align}
\hat{U}_{\rm kin}
=
\sum_i
\left(
\ket{i}\bra{i}
\otimes
\hat{I}
\right)
+
\sum_j
\left(
\ket{\tilde{j}}\bra{\tilde{j}}
\otimes\hat{\tilde{\Pi}}_j
\right)\, . 
\end{align}
This unitary is the block encoding of the bosonic Hamiltonian \eqref{eq:Hamiltonian-boson}. 

Now we introduce fermions. The only difference from the bosonic case is that $\hat{\Pi}_i$ in $\hat{U}_{\rm pot}$ is modified slightly due to fermions, including the auxiliary ones. Pauli strings become slightly longer (but the length does not grow with the lattice size $L$), and $X$ and $Y$ can also appear in addition to $Z$. 
%%%%%%%%%%%%%%%%%%%%%%
%%%%%%%%%%%%%%%%%%%%%%
\subsubsection{QCD via Verstraete--Cirac transform}\label{sec:gate_counting_QCD_VC}
%%%%%%%%%%%%%%%%%%%%%%
%%%%%%%%%%%%%%%%%%%%%%
We can promote the above argument to QCD, as we did for the Jordan--Wigner transform in Sec.~\ref{sec:gate_counting_QCD_JW}. 
In the case of the Jordan--Wigner transform, we needed to choose a specific ordering of the products in the Trotter decomposition to make CNOT gates cancel with each other. With the Verstraete--Cirac transform, however, we do not need a specific ordering. This leads to the reduction of the depth of the circuit and the possibility of parallelization.

Block encoding of the Hamiltonian as a linear combination of unitaries is also straightforward because the Pauli-string expansion takes a simple form. Specifically, the length of each Pauli string does not increase with $L$.

%%%%%%%%%%%%%%%%%%%%%%
%%%%%%%%%%%%%%%%%%%%%%
\section{Alternative fermion-to-qubit mappings}\label{sec:other_methods}
%%%%%%%%%%%%%%%%%%%%%%
%%%%%%%%%%%%%%%%%%%%%%
In the preceding sections, we examined the Jordan--Wigner and Verstraete--Cirac transforms, which are characterized by a direct mapping between individual fermionic operators, namely creation and annihilation operators, and structured strings of qubit Pauli operators. 

These mappings ensure that each operator preserves the canonical fermionic anticommutation relations within a qubit representation. However, alternative encoding strategies exist that can achieve more efficient mappings for specific classes of fermionic systems. One such method is the Bravyi–-Kitaev Superfast (BKSF) algorithm \cite{Bravyi2002AnPhy}, which encodes fermionic systems with geometrically local interactions using a graph framework.
In BKSF, fermionic modes are associated with the vertices of a graph, and 
the hopping operators between vertex-fermions are represented by qubit operators on the connecting edges. Qubits are assigned to the edges, and two classes of operators, one associated with each vertex and another with each edge, are defined to preserve the fermionic algebra. 
Crucially, the number of qubits generally differs from the number of fermionic modes, and the total Hilbert space of the qubit system is larger than that of the target fermionic system. To ensure that the qubit system reproduces the original fermionic dynamics faithfully, ``stabilizer constraints" are imposed. 

These constraints project the enlarged qubit Hilbert space onto the physical subspace corresponding to the original fermionic states. As we have seen, the penalty terms introduced in the Verstraete--Cirac framework serve a similar role. In that context, the ground state of the penalty terms (or stabilizer constraints) is a tensor product of physical degrees of freedom and the redundant auxiliary degrees of freedom. 
A similar decoupling occurs in the BKSF framework, where stabilizer conditions isolate the nonphysical degrees of freedom, ensuring that they remain disentangled from the encoded fermionic sector.

This perspective enables us to find more compact and collective encoding strategies. For example, one may directly map fermion bilinears, such as hopping terms and on-site fermion parity, to specific combinations of qubit operators. In such cases, the encoding captures only the algebraically and physically relevant operators and does not need to provide a standalone representation of each mode. This is the foundational insight behind recent developments in exact bosonization, where the focus lies on mapping the full algebra of physical observables while achieving reductions in qubit overhead and maintaining locality~\cite{Chen2023PRXQ}.

A natural figure of merit for evaluating the efficiency of fermion-to-qubit transformations is the number of ancilla qubits that used to eliminate the nonlocality, which was zero for Jordan--Wigner and $d$ per site for Verstraete--Cirac for $d$-dimensional lattice. More recent approaches -- particularly those based on collective encoding and stabilizer constraints -- allow a smaller number of ancillae. Such a reduction is possible by disentangling some qubits through finite-depth circuits and discarding them. In other words, fractional qubit-to-fermion ratio is possible because of collective encoding in which multiple fermions are encoded together in a stabilizer code subspace, not from partial qubit assignments to individual fermions.

The gate counts for quantum simulation using more compact fermion-to-qubit transformations mentioned above will be comparable to those of the Verstraete--Cirac transformation in the previous analysis, up to a constant factor.

%%%%%%%%%%%%%%%%%%%%%%
%%%%%%%%%%%%%%%%%%%%%%
\subsection*{Fermionic quantum computing}
%%%%%%%%%%%%%%%%%%%%%%
%%%%%%%%%%%%%%%%%%%%%%
Lastly, it is worth mentioning that fermionic degrees of freedom can be used as a replacement for qubits. Indeed, it was shown in Ref.~\cite{Bravyi2002AnPhy} that a universal gate set can be constructed from fermionic creation operator $\hat{a}^\dagger_{i}$ and annihilation operator $\hat{a}_i$, combining them into unitary operations as follows:
\begin{align}
U_1(\alpha) &= \exp\left(\mathrm{i}\alpha\, \hat{a}_i^\dagger \hat{a}_i\right), \nonumber\\
U_2(\beta) &= \exp\left(\mathrm{i}\beta\, (\hat{a}_i^\dagger \hat{a}_j + \hat{a}_j^\dagger \hat{a}_i)\right), \nonumber\\
U_3(\gamma) &= \exp\left(\mathrm{i}\gamma\, (\hat{a}_i^\dagger\hat{a}_j^\dagger + \hat{a}_j \hat{a}_i)\right), \nonumber\\
U_4(\delta) &= \exp\left(\mathrm{i}\delta\, \hat{a}_i^\dagger\hat{a}_i\hat{a}_j^\dagger \hat{a}_j\right).
\end{align}
A special example of a universal gate set is given by choosing $\alpha=\beta=\gamma=\frac{\pi}{4}$ and $\delta=\pi$.
Some possible routes toward experimental realization of such fermionic quantum computing architectures were proposed in Ref.~\cite{OBrien2018,Gonzalez2023,Vilkelis2024}.
By construction, fermionic quantum computing is well-suited for simulating fermionic systems. Specifically, there is no issue of nonlocal mappings to qubits.

%%%%%%%%%%%%%%%%%%%%%%
%%%%%%%%%%%%%%%%%%%%%%
\section{Conclusion and discussions}\label{sec:conclusion}
%%%%%%%%%%%%%%%%%%%%%%
%%%%%%%%%%%%%%%%%%%%%%
In this paper, we generalized a universal simulation protocol for bosonic systems~\cite{Halimeh:2024bth} to include fermions. To encode fermions, we used the Jordan--Wigner transform and the Verstraete--Cirac transform. These frameworks are applicable to QCD defined using the orbifold lattice formulation, and an exponential speedup compared to previous proposals can be achieved. 

When the Jordan--Wigner transform is applied to the Hamiltonian time evolution of a theory on an $L^d$ spatial lattice via Trotterization, one Trotter step can be realized using $\mathcal{O}(L^d)$ CNOT gates, Hadamard gates, phase gates, and one-qubit rotations. This is the case for any matter content and $\mathrm{SU}(N)$ gauge group at any $N$. While the Jordan--Wigner transform converts fermions to long Pauli strings, cancellation of quantum gates significantly simplifies the circuit. This scaling of the gate counting is already optimal --- it is proportional to the spatial volume, or equivalently, the number of terms in the lattice Hamiltonian. 
The same cancellation mechanism may also work for nonlocal interactions, depending on the details of the interactions. 
With the Verstraete--Cirac transform, Trotter time evolution and block encoding as a linear combination of unitaries can be realized efficiently. Our protocols do not assume a black box and hence are readily implementable on a fault-tolerant quantum computer. 

Kaplan, Katz, and \"{U}nsal~\cite{Kaplan:2002wv} invented the orbifold lattice to regularize supersymmetric Yang--Mills theory preserving a part of supersymmetry. Quantum simulation of supersymmetric Yang--Mills theory is of interest for many reasons, including rich dynamics and application to quantum gravity via holographic duality. The supersymmetric Hamiltonians invented in Ref.~\cite{Kaplan:2002wv} have some exotic structures, including fermions on diagonal links. It would be important to investigate whether the simulation protocols discussed in this paper can be generalized to them. 

The techniques discussed in this paper utilize local interactions and are not applicable to large-$N$ gauge theories including matrix models. If the Jordan--Wigner transform is applied to an SU($N$) theory at a large value of $N$, $\mathcal{O}(N^2)$ CNOT gates are required to describe a bilinear of adjoint fermions. Because there are $\mathcal{O}(N^3)$ terms in the fermionic part of the Hamiltonian and each of them requires $\mathcal{O}(N^2)$ CNOT gates, $\mathcal{O}(N^5)$ CNOT gates are required. On the other hand, there are $\mathcal{O}(N^4)$ terms in the bosonic part of the Hamiltonian, each of which requires $\mathcal{O}(N^0)$ gates. So, the fermionic part can be more costly. Although $N^5$ may not be that bad\footnote{
Note also that this scaling is not bad compared to the $N$ dependence of the cost of Markov Chain Monte Carlo simulation on a classical computer. If we naively compute the fermion determinant (the determinant of the Dirac operator), the cost is $\mathcal{O}(N^6)$ each time. By using the RHMC algorithm~\cite{Clark:2006fx} combined with the conjugate gradient algorithm and efficient implementation of multiplication of the Dirac operator to a vector utilizing the sparseness, we can reduce the cost so that the worst-case scaling is $\mathcal{O}(N^5)$. For details, see Refs.~\cite{hanada2022mcmc,Anagnostopoulos:2007fw,Catterall:2007fp}.} compared to $N^4$, it would be nice if we could find a good encoding of fermions that reduces the power of $N$. Such an efficient encoding, if it exists, would be particularly useful when we use matrix degrees of freedom to describe emergent spatial dimensions, so that nontrivial quantum field theories can be simulated on quantum computers~\cite{Gharibyan:2020bab}. Such a technique is useful to realize 4d $\mathcal{N}=4$ super Yang--Mills theory without parameter fine-tuning~\cite{Ishii:2008ib,Hanada:2010kt,Hanada:2010gs}. 

In this paper, we did not discuss state preparation. It is important to establish efficient state-preparation methods in the orbifold lattice formulation, utilizing its key advantage: both the coordinate and momentum bases can be used. 

In our universal framework, various quantum field theories are described as simple spin Hamiltonians with nonlocal interactions, up to a quantum Fourier transform that allows us to combine the kinetic and potential terms. Some models related to the SYK model, such as the spin-SYK model and the overlapping clusters SYK model, are good targets that share some essence of quantum simulation of quantum field theories, including QCD. We stress that, while we primarily have QCD in mind, our analysis applies to a rather generic class of quantum field theories. 

%%%%%%%%%%%%%%%%%%%%%%
%%%%%%%%%%%%%%%%%%%%%%
\begin{center}
\section*{Acknowledgments}
\end{center}
%%%%%%%%%%%%%%%%%%%%%%
%%%%%%%%%%%%%%%%%%%%%%
The authors thank Andreas Sch\"{a}fer for comments on the manuscript and Joe Salfi for valuable discussions.
M.H.~thanks the STFC for the support through the Consolidated Grant ST/Z001072/1 and the Royal Society for the International Exchanges award IEC/R3/213026. J.C.H.~acknowledges funding by the Max Planck Society, the Deutsche Forschungsgemeinschaft (DFG, German Research Foundation) under Germany’s Excellence Strategy – EXC-2111 – 390814868, and the European Research Council (ERC) under the European Union’s Horizon Europe research and innovation program (Grant Agreement No.~101165667) -- ERC Starting Grant QuSiGauge. This work is part of the Quantum Computing for High-Energy Physics (QC4HEP) working group.

\appendix
%%%%%%%%%%%%%%%%%%%%%%
%%%%%%%%%%%%%%%%%%%%%%
\section{Block encoding via Jordan--Wigner transform}\label{sec:block_encoding_JW}
%%%%%%%%%%%%%%%%%%%%%%
%%%%%%%%%%%%%%%%%%%%%%
In this appendix, we discuss a block encoding of the Hamiltonian of a boson-fermion system using the Jordan--Wigner transform. Unlike the case of Trotter decomposition, we do not find a simple cancellation mechanism for CNOT gates. This motivates us to consider other approaches; see Sec.~\ref{sec:Verstraete-Cirac}. 

The only difference from Sec.~\ref{sec:universal_framework_VC} is that $\hat{\Pi}_i$ in $\hat{U}_{\rm pot}$ can be a long Pauli string. 
In the case of the Trotter decomposition, we converted a long chain of $Z$ gates to a chain of CNOT gates. For theories with local interactions, we observed a cancellation of many CNOT gates coming from products of two chains. For block encoding, however, we do not find the same cancellation mechanism, because the chain of CNOT gates appears as a sum and not as a product. Therefore, we cannot reduce the number of gates relying on a cancellation; some $\hat{\Pi}$s require $\mathcal{O}(L^d)$ CNOT gates. Still, the depth can be suppressed to $\log L$, using a technique discussed in Sec.~\ref{sec:rearrangement_of_product}. Specifically, we can use
\begin{align}
Z_1\otimes\cdots\otimes Z_{2^\ell}
    &=
    \left(
    \prod_{j=1}^{2^{\ell-1}}C_{2j-1,2j}
    \right)   
    \left(
    \prod_{j=1}^{2^{\ell-2}}C_{4j-2,4j}
    \right)
    \cdots
    \left(
    \prod_{j=1}^{2}C_{2^{\ell-1}j-2^{\ell-2},2^{\ell-1}j}
    \right)  
    C_{2^{\ell-1},2^\ell}
    \nonumber\\
    &\qquad
    \times Z_{2^\ell}
    \nonumber\\
    &\qquad
    \times C_{2^{\ell-1},2^\ell}
    \left(
    \prod_{j=1}^{2}C_{2^{\ell-1}j-2^{\ell-2},2^{\ell-1}j}
    \right)  
    \cdots
    \left(
    \prod_{j=1}^{2^{\ell-2}}C_{4j-2,4j}
    \right)
        \left(
    \prod_{j=1}^{2^{\ell-1}}C_{2j-1,2j}
    \right)\, .    
\end{align}
%%%%%%%%%%%%%%%%%%%%%%
%%%%%%%%%%%%%%%%%%%%%%
\section{Trotter decomposition via Jordan--Wigner transform with a different ordering of lattice points}\label{sec:JW_another_ordering}
%%%%%%%%%%%%%%%%%%%%%%
%%%%%%%%%%%%%%%%%%%%%%
In this appendix, we apply the Jordan--Wigner transform with the ordering of the points used in Sec.~\ref{sec:Verstraete-Cirac} for the Verstraete--Cirac transform. Having QCD in mind, we assume fermion terms of the form $\hat{O}_{n,n'}\hat{\Psi}^{(\dagger)}_{n}\hat{\Psi}^{(\dagger)}_{n'}$, where $n$ and $n'$ are neighboring lattice points. $\hat{O}_{n,n'}$ depends only on bosons. (Here, we ignore `internal' degrees of freedom at each point and assume there is one complex fermion at each point. Generalization to QCD is straightforward, as we did in Sec.~\ref{sec:universal_framework_boson+fermion}.)

Below, we consider the case of $d=2$. (The generalization to $d>2$ is straightforward.) Then, the lattice points are ordered as in Fig.~\ref{fig:VC-2d}. For simplicity, let us assume the open boundary condition. (We will discuss the generalization to the periodic boundary condition later.)
Then, nontrivial long Pauli strings arise from the interaction along the second direction, i.e., when $n$ and $n'$ correspond to $(n_1,n_2)$ and $(n_1,n_2+1)$. For $n_2=1$, and $n_1=1,2,\cdots,L-1$, we have the following interaction term after applying the Jordan--Wigner transform:
\begin{align}
    V_{n_1}
    \equiv
    \sigma_{n_1}\otimes Z_{n_1+1}\otimes\cdots\otimes Z_{2L-n_1}\otimes\sigma_{2L+1-n_1}
    \otimes\hat{O}_{n_1}(\hat{x})\, . 
\end{align}
For $n_2=1$ and $n_1=L$, 
\begin{align}
    V_{n_1}
    \equiv
    \sigma_{L}\otimes\sigma_{L+1}
    \otimes\hat{O}_{n_1}(\hat{x})\, . 
\end{align}

For $n_1=1,2,\cdots,L-1$, we can write it as
\begin{align}
    V_{n_1}
    =
    \mathcal{C}_{n_1}
    \cdot
    \sigma_{n_1}\otimes Z_{L-1}\otimes\sigma_{2L+1-n_1}
    \otimes\hat{O}_{n_1}(\hat{x})
    \cdot
    \mathcal{C}_{n_1}\, , 
\end{align}
where
\begin{align}
    \mathcal{C}_{n_1}
    =
    C_{n_1+1,L-1}
   \cdots
   C_{L-2,L-1}
   \cdot
   C_{L,L-1}
   \cdots
   C_{2L-n_1,L-1}\, . 
\end{align}
Hence, the Trotter time evolution can be written as a product of 
\begin{align}
    \exp\left(-\mathrm{i}\epsilon V_{n_1}\right)
    =
    \mathcal{C}_{n_1}
    \cdot
    \exp\left(-\mathrm{i}\epsilon\sigma_{n_1}\otimes Z_{L-1}\otimes\sigma_{2L+1-n_1}
    \otimes\hat{O}_{n_1}(\hat{x})
    \right)
    \cdot
    \mathcal{C}_{n_1}\, .  
\end{align}
We can see a significant cancellation of CNOT gates because of 
\begin{align}
    \mathcal{C}_{n_1}
    \mathcal{C}_{n_1+1}
    =
    C_{n_1+1,L-1}
    \cdot    
    C_{2L-n_1,L-1}\, . 
    \label{eq:C_{n_1}}
\end{align}
The same cancellation can be found for $n_2=2,3,\cdots,L-1$, too. 

When the boundary condition is periodic, we need to take into account blue and orange links in Fig.~\ref{fig:2d-lattice}. 

The same cancellation mechanism as above applies to the orange links. There is a small difference for odd $L$ and even $L$; below, let us assume $L$ is even, where the math becomes slightly simpler. (Note that Fig.~\ref{fig:VC-2d} is for odd $L$.)
Then, For $n_1=1,2,\cdots,L$, we have the following interaction term after applying the Jordan--Wigner transform:
\begin{align}
    V'_{n_1}
    \equiv
    \sigma_{n_1}\otimes Z_{n_1+1}\otimes\cdots\otimes Z_{L^2-n_1}\otimes\sigma_{L^2+1-n_1}
    \otimes\hat{O}'_{n_1}(\hat{x})\, . 
\end{align}
Therefore, exactly the same cancellation mechanism as above works, just replacing $2L$ with $L^2$. 

From blue links, we obtain
\begin{align}
    V''_{n_2}
    \equiv
    \sigma_{(n_2-1)L+1}\otimes Z_{(n_2-1)L+2}\otimes\cdots\otimes Z_{n_2L-1}\otimes\sigma_{n_2L}
    \otimes\hat{O}''_{n_2}(\hat{x}) 
\end{align}
and
\begin{align}
    \exp\left(-\mathrm{i}\epsilon V''_{n_2}\right)
    =
\mathcal{C}'_{n_2}
    \cdot
    \exp\left(-\mathrm{i}\epsilon
        \sigma_{(n_2-1)L+1}\otimes
        Z_{n_2L-1}\otimes\sigma_{n_2L}
    \otimes\hat{O}''_{n_2}(\hat{x})
    \right)
    \cdot
    \mathcal{C}'_{n_2}\, , 
\end{align}
where
\begin{align}
 \mathcal{C}'_{n_2}
 =
    \prod_{j={(n_2-1)L+2}}^{n_2L-2}
    C_{j,n_2L-1}\, . 
\end{align}
This $\mathcal{C}'_{n_2}$ can be canceled with a part of the CNOT gates that remain uncanceled in the above. 

\bibliographystyle{utphys}
\bibliography{biblio,reference}

\providecommand{\href}[2]{#2}\begingroup\raggedright\begin{thebibliography}{100}

\bibitem{Halimeh:2024bth}
J.~C. Halimeh, M.~Hanada, S.~Matsuura, F.~Nori, E.~Rinaldi, and A.~Sch\"afer,
  ``{A universal framework for the quantum simulation of Yang-Mills theory},''
  \href{http://arxiv.org/abs/2411.13161}{{\ttfamily arXiv:2411.13161
  [quant-ph]}}.

\bibitem{Dalmonte:2016alw}
M.~Dalmonte and S.~Montangero, ``{Lattice gauge theory simulations in the
  quantum information era},''
  \href{http://dx.doi.org/10.1080/00107514.2016.1151199}{{\em Contemp. Phys.}
  {\bfseries 57} no.~3, (2016) 388--412},
  \href{http://arxiv.org/abs/1602.03776}{{\ttfamily arXiv:1602.03776
  [cond-mat.quant-gas]}}.

\bibitem{Banuls:2019bmf}
M.~C. Ba\~nuls {\em et~al.}, ``{Simulating Lattice Gauge Theories within
  Quantum Technologies},''
  \href{http://dx.doi.org/10.1140/epjd/e2020-100571-8}{{\em Eur. Phys. J. D}
  {\bfseries 74} no.~8, (2020) 165},
  \href{http://arxiv.org/abs/1911.00003}{{\ttfamily arXiv:1911.00003
  [quant-ph]}}.

\bibitem{Zohar_review}
E.~Zohar, J.~I. Cirac, and B.~Reznik, ``Quantum simulations of lattice gauge
  theories using ultracold atoms in optical lattices,''
  \href{http://dx.doi.org/10.1088/0034-4885/79/1/014401}{{\em Rep. Prog. Phys.}
  {\bfseries 79} no.~1, (Dec, 2015) 014401}.

\bibitem{Aidelsburger:2021mia}
M.~Aidelsburger {\em et~al.}, ``{Cold atoms meet lattice gauge theory},''
  \href{http://dx.doi.org/10.1098/rsta.2021.0064}{{\em Phil. Trans. Roy. Soc.
  Lond. A} {\bfseries 380} (2021) 20210064},
  \href{http://arxiv.org/abs/2106.03063}{{\ttfamily arXiv:2106.03063
  [cond-mat.quant-gas]}}.

\bibitem{Zohar_NewReview}
E.~{Zohar}, ``{Quantum simulation of lattice gauge theories in more than one
  space dimension{\textemdash}requirements, challenges and methods},''
  \href{http://dx.doi.org/10.1098/rsta.2021.0069}{{\em Philos. Trans. Royal
  Soc. A} {\bfseries 380} no.~2216, (Feb., 2022) 20210069},
  \href{http://arxiv.org/abs/2106.04609}{{\ttfamily arXiv:2106.04609
  [quant-ph]}}.

\bibitem{klco2021standard}
N.~Klco, A.~Roggero, and M.~J. Savage, ``Standard model physics and the digital
  quantum revolution: thoughts about the interface,''
  \href{http://dx.doi.org/10.1088/1361-6633/ac58a4}{{\em Reports on Progress in
  Physics} {\bfseries 85} no.~6, (May, 2022) 064301}.
  \url{https://dx.doi.org/10.1088/1361-6633/ac58a4}.

\bibitem{Bauer:2022hpo}
C.~W. Bauer {\em et~al.}, ``{Quantum Simulation for High-Energy Physics},''
  \href{http://dx.doi.org/10.1103/PRXQuantum.4.027001}{{\em PRX Quantum}
  {\bfseries 4} no.~2, (2023) 027001},
  \href{http://arxiv.org/abs/2204.03381}{{\ttfamily arXiv:2204.03381
  [quant-ph]}}.

\bibitem{DiMeglio:2023nsa}
A.~Di~Meglio {\em et~al.}, ``{Quantum Computing for High-Energy Physics: State
  of the Art and Challenges},''
  \href{http://dx.doi.org/10.1103/PRXQuantum.5.037001}{{\em PRX Quantum}
  {\bfseries 5} no.~3, (2024) 037001},
  \href{http://arxiv.org/abs/2307.03236}{{\ttfamily arXiv:2307.03236
  [quant-ph]}}.

\bibitem{Cheng_review}
Y.~Cheng and H.~Zhai, ``Emergent u(1) lattice gauge theory in rydberg atom
  arrays,'' \href{http://dx.doi.org/10.1038/s42254-024-00749-6}{{\em Nature
  Reviews Physics} {\bfseries 6} no.~9, (2024) 566--576}.
  \url{https://doi.org/10.1038/s42254-024-00749-6}.

\bibitem{Halimeh_review}
J.~C. Halimeh, M.~Aidelsburger, F.~Grusdt, P.~Hauke, and B.~Yang, ``Cold-atom
  quantum simulators of gauge theories,''
  \href{http://dx.doi.org/10.1038/s41567-024-02721-8}{{\em Nature Physics}
  (2025) }. \url{https://doi.org/10.1038/s41567-024-02721-8}.

\bibitem{Cohen:2021imf}
{\bfseries NuQS} Collaboration, T.~D. Cohen, H.~Lamm, S.~Lawrence, and
  Y.~Yamauchi, ``{Quantum algorithms for transport coefficients in gauge
  theories},'' \href{http://dx.doi.org/10.1103/PhysRevD.104.094514}{{\em Phys.
  Rev. D} {\bfseries 104} no.~9, (2021) 094514},
  \href{http://arxiv.org/abs/2104.02024}{{\ttfamily arXiv:2104.02024
  [hep-lat]}}.

\bibitem{Lee:2024jnt}
K.~Lee, F.~Turro, and X.~Yao, ``Quantum computing for energy correlators,''
  \href{http://dx.doi.org/10.1103/PhysRevD.111.054514}{{\em Phys. Rev. D}
  {\bfseries 111} (Mar, 2025) 054514}.
  \url{https://link.aps.org/doi/10.1103/PhysRevD.111.054514}.

\bibitem{Turro:2024pxu}
F.~Turro, A.~Ciavarella, and X.~Yao, ``{Classical and quantum computing of
  shear viscosity for (2+1)D SU(2) gauge theory},''
  \href{http://dx.doi.org/10.1103/PhysRevD.109.114511}{{\em Phys. Rev. D}
  {\bfseries 109} no.~11, (2024) 114511},
  \href{http://arxiv.org/abs/2402.04221}{{\ttfamily arXiv:2402.04221
  [hep-lat]}}.

\bibitem{Zohar:2011cw}
E.~Zohar and B.~Reznik, ``{Confinement and lattice QED electric flux-tubes
  simulated with ultracold atoms},''
  \href{http://dx.doi.org/10.1103/PhysRevLett.107.275301}{{\em Phys. Rev.
  Lett.} {\bfseries 107} (2011) 275301},
  \href{http://arxiv.org/abs/1108.1562}{{\ttfamily arXiv:1108.1562
  [quant-ph]}}.

\bibitem{Zohar:2012ay}
E.~Zohar, J.~I. Cirac, and B.~Reznik, ``{Simulating Compact Quantum
  Electrodynamics with ultracold atoms: Probing confinement and nonperturbative
  effects},'' \href{http://dx.doi.org/10.1103/PhysRevLett.109.125302}{{\em
  Phys. Rev. Lett.} {\bfseries 109} (2012) 125302},
  \href{http://arxiv.org/abs/1204.6574}{{\ttfamily arXiv:1204.6574
  [quant-ph]}}.

\bibitem{Banerjee2012}
D.~Banerjee, M.~Dalmonte, M.~Müller, E.~Rico, P.~Stebler, U.-J. Wiese, and
  P.~Zoller, ``Atomic quantum simulation of dynamical gauge fields coupled to
  fermionic matter: From string breaking to evolution after a quench,''
  \href{http://dx.doi.org/10.1103/physrevlett.109.175302}{{\em Phys. Rev.
  Lett.} {\bfseries 109} no.~17, (Oct, 2012) }.
  \url{http://dx.doi.org/10.1103/PhysRevLett.109.175302}.

\bibitem{Zohar2013}
E.~Zohar, J.~I. Cirac, and B.~Reznik, ``Simulating ($2+1$)-dimensional lattice
  qed with dynamical matter using ultracold atoms,''
  \href{http://dx.doi.org/10.1103/PhysRevLett.110.055302}{{\em Phys. Rev.
  Lett.} {\bfseries 110} (Jan, 2013) 055302}.
  \url{https://link.aps.org/doi/10.1103/PhysRevLett.110.055302}.

\bibitem{Banerjee2013atomic}
D.~Banerjee, M.~B\"ogli, M.~Dalmonte, E.~Rico, P.~Stebler, U.-J. Wiese, and
  P.~Zoller, ``Atomic quantum simulation of $\mathbf{U}(n)$ and
  $\mathrm{SU}(n)$ non-abelian lattice gauge theories,''
  \href{http://dx.doi.org/10.1103/PhysRevLett.110.125303}{{\em Phys. Rev.
  Lett.} {\bfseries 110} (Mar, 2013) 125303}.
  \url{https://link.aps.org/doi/10.1103/PhysRevLett.110.125303}.

\bibitem{Halimeh2022tuning}
J.~C. Halimeh, I.~P. McCulloch, B.~Yang, and P.~Hauke, ``Tuning the topological
  $\ensuremath{\theta}$-angle in cold-atom quantum simulators of gauge
  theories,'' \href{http://dx.doi.org/10.1103/PRXQuantum.3.040316}{{\em PRX
  Quantum} {\bfseries 3} (Nov, 2022) 040316}.
  \url{https://link.aps.org/doi/10.1103/PRXQuantum.3.040316}.

\bibitem{Cheng2022}
Y.~Cheng, S.~Liu, W.~Zheng, P.~Zhang, and H.~Zhai, ``Tunable
  confinement-deconfinement transition in an ultracold-atom quantum
  simulator,'' \href{http://dx.doi.org/10.1103/PRXQuantum.3.040317}{{\em PRX
  Quantum} {\bfseries 3} (Nov, 2022) 040317}.
  \url{https://link.aps.org/doi/10.1103/PRXQuantum.3.040317}.

\bibitem{Fontana2022}
P.~Fontana, J.~C.~P. Barros, and A.~Trombettoni, ``Quantum simulator of link
  models using spinor dipolar ultracold atoms,''.
  \url{https://arxiv.org/abs/2210.14836}.

\bibitem{surace2023abinitio}
F.~M. Surace, P.~Fromholz, N.~D. Oppong, M.~Dalmonte, and M.~Aidelsburger,
  ``$ab\,initio$ derivation of lattice gauge theory dynamics for cold gases in
  optical lattices,'' \href{http://dx.doi.org/10.1103/PRXQuantum.4.020330}{{\em
  PRX Quantum} {\bfseries 4} (2023) 020330}.
  \url{https://link.aps.org/doi/10.1103/PRXQuantum.4.020330}.

\bibitem{osborne2022largescale}
J.~Osborne, I.~P. McCulloch, B.~Yang, P.~Hauke, and J.~C. Halimeh,
  ``Large-scale $2+1$d $\mathrm{U}(1)$ gauge theory with dynamical matter in a
  cold-atom quantum simulator,''
  \href{http://arxiv.org/abs/2211.01380}{{\ttfamily arXiv:2211.01380
  [cond-mat.quant-gas]}}.

\bibitem{osborne2023spins}
J.~Osborne, B.~Yang, I.~P. McCulloch, P.~Hauke, and J.~C. Halimeh, ``Spin-$s$
  $\mathrm{U}(1)$ quantum link models with dynamical matter on a quantum
  simulator,'' \href{http://arxiv.org/abs/2305.06368}{{\ttfamily
  arXiv:2305.06368 [cond-mat.quant-gas]}}.
  \url{https://arxiv.org/abs/2305.06368}.

\bibitem{Martinez2016}
E.~A. Martinez, C.~A. Muschik, P.~Schindler, D.~Nigg, A.~Erhard, M.~Heyl,
  P.~Hauke, M.~Dalmonte, T.~Monz, P.~Zoller, and R.~Blatt, ``Real-time dynamics
  of lattice gauge theories with a few-qubit quantum computer,''
  \href{http://dx.doi.org/10.1038/nature18318}{{\em Nature} {\bfseries 534}
  no.~7608, (2016) 516--519}. \url{https://doi.org/10.1038/nature18318}.

\bibitem{Klco2018}
N.~Klco, E.~F. Dumitrescu, A.~J. McCaskey, T.~D. Morris, R.~C. Pooser, M.~Sanz,
  E.~Solano, P.~Lougovski, and M.~J. Savage, ``Quantum-classical computation of
  {Schwinger} model dynamics using quantum computers,''
  \href{http://dx.doi.org/10.1103/PhysRevA.98.032331}{{\em Phys. Rev. A}
  {\bfseries 98} (Sep, 2018) 032331}.
  \url{https://link.aps.org/doi/10.1103/PhysRevA.98.032331}.

\bibitem{Goerg2019}
F.~G{\"o}rg, K.~Sandholzer, J.~Minguzzi, R.~Desbuquois, M.~Messer, and
  T.~Esslinger, ``Realization of density-dependent {Peierls} phases to engineer
  quantized gauge fields coupled to ultracold matter,''
  \href{http://dx.doi.org/10.1038/s41567-019-0615-4}{{\em Nat. Phys.}
  {\bfseries 15} no.~11, (2019) 1161--1167}.
  \url{https://doi.org/10.1038/s41567-019-0615-4}.

\bibitem{Schweizer2019}
C.~Schweizer, F.~Grusdt, M.~Berngruber, L.~Barbiero, E.~Demler, N.~Goldman,
  I.~Bloch, and M.~Aidelsburger, ``Floquet approach to $\mathbb{Z}$2 lattice
  gauge theories with ultracold atoms in optical lattices,''
  \href{http://dx.doi.org/10.1038/s41567-019-0649-7}{{\em Nat. Phys.}
  {\bfseries 15} no.~11, (2019) 1168--1173}.
  \url{https://doi.org/10.1038/s41567-019-0649-7}.

\bibitem{Mil2020}
A.~Mil, T.~V. Zache, A.~Hegde, A.~Xia, R.~P. Bhatt, M.~K. Oberthaler, P.~Hauke,
  J.~Berges, and F.~Jendrzejewski, ``A scalable realization of local {U(1)}
  gauge invariance in cold atomic mixtures,''
  \href{http://dx.doi.org/10.1126/science.aaz5312}{{\em Science} {\bfseries
  367} no.~6482, (2020) 1128--1130}.
  \url{https://science.sciencemag.org/content/367/6482/1128}.

\bibitem{Yang2020}
B.~Yang, H.~Sun, R.~Ott, H.-Y. Wang, T.~V. Zache, J.~C. Halimeh, Z.-S. Yuan,
  P.~Hauke, and J.-W. Pan, ``Observation of gauge invariance in a 71-site
  {Bose--Hubbard} quantum simulator,''
  \href{http://dx.doi.org/10.1038/s41586-020-2910-8}{{\em Nature} {\bfseries
  587} no.~7834, (2020) 392--396}.
  \url{https://doi.org/10.1038/s41586-020-2910-8}.

\bibitem{Wang2021}
Z.~Wang, Z.-Y. Ge, Z.~Xiang, X.~Song, R.-Z. Huang, P.~Song, X.-Y. Guo, L.~Su,
  K.~Xu, D.~Zheng, and H.~Fan, ``Observation of emergent ${\mathbb{z}}_{2}$
  gauge invariance in a superconducting circuit,''
  \href{http://dx.doi.org/10.1103/PhysRevResearch.4.L022060}{{\em Phys. Rev.
  Research} {\bfseries 4} (Jun, 2022) L022060}.
  \url{https://link.aps.org/doi/10.1103/PhysRevResearch.4.L022060}.

\bibitem{Zhou2022}
Z.-Y. Zhou, G.-X. Su, J.~C. Halimeh, R.~Ott, H.~Sun, P.~Hauke, B.~Yang, Z.-S.
  Yuan, J.~Berges, and J.-W. Pan, ``Thermalization dynamics of a gauge theory
  on a quantum simulator,''
  \href{http://dx.doi.org/10.1126/science.abl6277}{{\em Science} {\bfseries
  377} no.~6603, (2022) 311--314}.

\bibitem{Wang2023}
H.-Y. Wang, W.-Y. Zhang, Z.~Yao, Y.~Liu, Z.-H. Zhu, Y.-G. Zheng, X.-K. Wang,
  H.~Zhai, Z.-S. Yuan, and J.-W. Pan, ``Interrelated thermalization and quantum
  criticality in a lattice gauge simulator,''
  \href{http://dx.doi.org/10.1103/PhysRevLett.131.050401}{{\em Phys. Rev.
  Lett.} {\bfseries 131} (Aug, 2023) 050401}.
  \url{https://link.aps.org/doi/10.1103/PhysRevLett.131.050401}.

\bibitem{Zhang:2023hzr}
W.-Y. Zhang {\em et~al.}, ``{Observation of microscopic confinement dynamics by
  a tunable topological \ensuremath{\theta}-angle},''
  \href{http://dx.doi.org/10.1038/s41567-024-02702-x}{{\em Nature Phys.}
  {\bfseries 21} no.~1, (2025) 155--160},
  \href{http://arxiv.org/abs/2306.11794}{{\ttfamily arXiv:2306.11794
  [cond-mat.quant-gas]}}.

\bibitem{Ciavarella2024quantum}
A.~N. Ciavarella and C.~W. Bauer, ``Quantum simulation of su(3) lattice
  yang-mills theory at leading order in large-${N}_{c}$ expansion,''
  \href{http://dx.doi.org/10.1103/PhysRevLett.133.111901}{{\em Phys. Rev.
  Lett.} {\bfseries 133} (Sep, 2024) 111901}.
  \url{https://link.aps.org/doi/10.1103/PhysRevLett.133.111901}.

\bibitem{Ciavarella:2024lsp}
A.~N. Ciavarella, ``{String Breaking in the Heavy Quark Limit with Scalable
  Circuits},'' \href{http://arxiv.org/abs/2411.05915}{{\ttfamily
  arXiv:2411.05915 [quant-ph]}}.

\bibitem{de2024observationstringbreakingdynamicsquantum}
A.~De, A.~Lerose, D.~Luo, F.~M. Surace, A.~Schuckert, E.~R. Bennewitz, B.~Ware,
  W.~Morong, K.~S. Collins, Z.~Davoudi, A.~V. Gorshkov, O.~Katz, and C.~Monroe,
  ``Observation of string-breaking dynamics in a quantum simulator,''
  \href{http://arxiv.org/abs/2410.13815}{{\ttfamily arXiv:2410.13815
  [quant-ph]}}. \url{https://arxiv.org/abs/2410.13815}.

\bibitem{liu2024stringbreakingmechanismlattice}
Y.~Liu, W.-Y. Zhang, Z.-H. Zhu, M.-G. He, Z.-S. Yuan, and J.-W. Pan, ``String
  breaking mechanism in a lattice schwinger model simulator,''
  \href{http://arxiv.org/abs/2411.15443}{{\ttfamily arXiv:2411.15443
  [cond-mat.quant-gas]}}. \url{https://arxiv.org/abs/2411.15443}.

\bibitem{Farrell:2023fgd}
R.~C. Farrell, M.~Illa, A.~N. Ciavarella, and M.~J. Savage, ``{Scalable
  Circuits for Preparing Ground States on Digital Quantum Computers: The
  Schwinger Model Vacuum on 100 Qubits},''
  \href{http://dx.doi.org/10.1103/PRXQuantum.5.020315}{{\em PRX Quantum}
  {\bfseries 5} no.~2, (2024) 020315},
  \href{http://arxiv.org/abs/2308.04481}{{\ttfamily arXiv:2308.04481
  [quant-ph]}}.

\bibitem{Farrell:2024fit}
R.~C. Farrell, M.~Illa, A.~N. Ciavarella, and M.~J. Savage, ``{Quantum
  simulations of hadron dynamics in the Schwinger model using 112 qubits},''
  \href{http://dx.doi.org/10.1103/PhysRevD.109.114510}{{\em Phys. Rev. D}
  {\bfseries 109} no.~11, (2024) 114510},
  \href{http://arxiv.org/abs/2401.08044}{{\ttfamily arXiv:2401.08044
  [quant-ph]}}.

\bibitem{zhu2024probingfalsevacuumdecay}
Z.-H. Zhu, Y.~Liu, G.~Lagnese, F.~M. Surace, W.-Y. Zhang, M.-G. He, J.~C.
  Halimeh, M.~Dalmonte, S.~C. Morampudi, F.~Wilczek, Z.-S. Yuan, and J.-W. Pan,
  ``Probing false vacuum decay on a cold-atom gauge-theory quantum simulator,''
  \href{http://arxiv.org/abs/2411.12565}{{\ttfamily arXiv:2411.12565
  [cond-mat.quant-gas]}}. \url{https://arxiv.org/abs/2411.12565}.

\bibitem{Ciavarella:2021nmj}
A.~Ciavarella, N.~Klco, and M.~J. Savage, ``{Trailhead for quantum simulation
  of SU(3) Yang-Mills lattice gauge theory in the local multiplet basis},''
  \href{http://dx.doi.org/10.1103/PhysRevD.103.094501}{{\em Phys. Rev. D}
  {\bfseries 103} no.~9, (2021) 094501},
  \href{http://arxiv.org/abs/2101.10227}{{\ttfamily arXiv:2101.10227
  [quant-ph]}}.

\bibitem{Ciavarella:2023mfc}
A.~N. Ciavarella, ``{Quantum simulation of lattice QCD with improved
  Hamiltonians},'' \href{http://dx.doi.org/10.1103/PhysRevD.108.094513}{{\em
  Phys. Rev. D} {\bfseries 108} no.~9, (2023) 094513},
  \href{http://arxiv.org/abs/2307.05593}{{\ttfamily arXiv:2307.05593
  [hep-lat]}}.

\bibitem{Ciavarella:2021lel}
A.~N. Ciavarella and I.~A. Chernyshev, ``{Preparation of the SU(3) lattice
  Yang-Mills vacuum with variational quantum methods},''
  \href{http://dx.doi.org/10.1103/PhysRevD.105.074504}{{\em Phys. Rev. D}
  {\bfseries 105} no.~7, (2022) 074504},
  \href{http://arxiv.org/abs/2112.09083}{{\ttfamily arXiv:2112.09083
  [quant-ph]}}.

\bibitem{Gustafson:2023kvd}
E.~J. Gustafson, H.~Lamm, and F.~Lovelace, ``{Primitive quantum gates for an
  SU(2) discrete subgroup: Binary octahedral},''
  \href{http://dx.doi.org/10.1103/PhysRevD.109.054503}{{\em Phys. Rev. D}
  {\bfseries 109} no.~5, (2024) 054503},
  \href{http://arxiv.org/abs/2312.10285}{{\ttfamily arXiv:2312.10285
  [hep-lat]}}.

\bibitem{Gustafson:2024kym}
E.~J. Gustafson, Y.~Ji, H.~Lamm, E.~M. Murairi, S.~O. Perez, and S.~Zhu,
  ``{Primitive quantum gates for an SU(3) discrete subgroup:
  \ensuremath{\Sigma}(36\texttimes{}3)},''
  \href{http://dx.doi.org/10.1103/PhysRevD.110.034515}{{\em Phys. Rev. D}
  {\bfseries 110} no.~3, (2024) 034515},
  \href{http://arxiv.org/abs/2405.05973}{{\ttfamily arXiv:2405.05973
  [hep-lat]}}.

\bibitem{Lamm:2024jnl}
H.~Lamm, Y.-Y. Li, J.~Shu, Y.-L. Wang, and B.~Xu, ``{Block encodings of
  discrete subgroups on a quantum computer},''
  \href{http://dx.doi.org/10.1103/PhysRevD.110.054505}{{\em Phys. Rev. D}
  {\bfseries 110} no.~5, (2024) 054505},
  \href{http://arxiv.org/abs/2405.12890}{{\ttfamily arXiv:2405.12890
  [hep-lat]}}.

\bibitem{Farrell:2022wyt}
R.~C. Farrell, I.~A. Chernyshev, S.~J.~M. Powell, N.~A. Zemlevskiy, M.~Illa,
  and M.~J. Savage, ``{Preparations for quantum simulations of quantum
  chromodynamics in 1+1 dimensions. I. Axial gauge},''
  \href{http://dx.doi.org/10.1103/PhysRevD.107.054512}{{\em Phys. Rev. D}
  {\bfseries 107} no.~5, (2023) 054512},
  \href{http://arxiv.org/abs/2207.01731}{{\ttfamily arXiv:2207.01731
  [quant-ph]}}.

\bibitem{Farrell:2022vyh}
R.~C. Farrell, I.~A. Chernyshev, S.~J.~M. Powell, N.~A. Zemlevskiy, M.~Illa,
  and M.~J. Savage, ``{Preparations for quantum simulations of quantum
  chromodynamics in 1+1 dimensions. II. Single-baryon \ensuremath{\beta}-decay
  in real time},'' \href{http://dx.doi.org/10.1103/PhysRevD.107.054513}{{\em
  Phys. Rev. D} {\bfseries 107} no.~5, (2023) 054513},
  \href{http://arxiv.org/abs/2209.10781}{{\ttfamily arXiv:2209.10781
  [quant-ph]}}.

\bibitem{Li:2024lrl}
Z.~Li, D.~M. Grabowska, and M.~J. Savage, ``{Sequency Hierarchy Truncation
  (SeqHT) for Adiabatic State Preparation and Time Evolution in Quantum
  Simulations},'' \href{http://arxiv.org/abs/2407.13835}{{\ttfamily
  arXiv:2407.13835 [quant-ph]}}.

\bibitem{Zemlevskiy:2024vxt}
N.~A. Zemlevskiy, ``{Scalable Quantum Simulations of Scattering in Scalar Field
  Theory on 120 Qubits},'' \href{http://arxiv.org/abs/2411.02486}{{\ttfamily
  arXiv:2411.02486 [quant-ph]}}.

\bibitem{Lewis:2019wfx}
R.~Lewis and R.~M. Woloshyn, ``{A qubit model for U(1) lattice gauge theory},''
\newblock 5, 2019.
\newblock \href{http://arxiv.org/abs/1905.09789}{{\ttfamily arXiv:1905.09789
  [hep-lat]}}.

\bibitem{Atas:2021ext}
Y.~Y. Atas, J.~Zhang, R.~Lewis, A.~Jahanpour, J.~F. Haase, and C.~A. Muschik,
  ``{SU(2) hadrons on a quantum computer via a variational approach},''
  \href{http://dx.doi.org/10.1038/s41467-021-26825-4}{{\em Nature Commun.}
  {\bfseries 12} no.~1, (2021) 6499},
  \href{http://arxiv.org/abs/2102.08920}{{\ttfamily arXiv:2102.08920
  [quant-ph]}}.

\bibitem{ARahman:2022tkr}
S.~A~Rahman, R.~Lewis, E.~Mendicelli, and S.~Powell, ``{Self-mitigating Trotter
  circuits for SU(2) lattice gauge theory on a quantum computer},''
  \href{http://dx.doi.org/10.1103/PhysRevD.106.074502}{{\em Phys. Rev. D}
  {\bfseries 106} no.~7, (2022) 074502},
  \href{http://arxiv.org/abs/2205.09247}{{\ttfamily arXiv:2205.09247
  [hep-lat]}}.

\bibitem{Atas:2022dqm}
Y.~Y. Atas, J.~F. Haase, J.~Zhang, V.~Wei, S.~M.~L. Pfaendler, R.~Lewis, and
  C.~A. Muschik, ``{Simulating one-dimensional quantum chromodynamics on a
  quantum computer: Real-time evolutions of tetra- and pentaquarks},''
  \href{http://dx.doi.org/10.1103/PhysRevResearch.5.033184}{{\em Phys. Rev.
  Res.} {\bfseries 5} no.~3, (2023) 033184},
  \href{http://arxiv.org/abs/2207.03473}{{\ttfamily arXiv:2207.03473
  [quant-ph]}}.

\bibitem{Mendicelli:2022ntz}
E.~Mendicelli, R.~Lewis, S.~A. Rahman, and S.~Powell, ``{Real time evolution
  and a traveling excitation in SU(2) pure gauge theory on a quantum
  computer.},'' \href{http://dx.doi.org/10.22323/1.430.0025}{{\em PoS}
  {\bfseries LATTICE2022} (2023) 025},
  \href{http://arxiv.org/abs/2210.11606}{{\ttfamily arXiv:2210.11606
  [hep-lat]}}.

\bibitem{Kavaki:2024ijd}
A.~H.~Z. Kavaki and R.~Lewis, ``{From square plaquettes to triamond lattices
  for SU(2) gauge theory},''
  \href{http://dx.doi.org/10.1038/s42005-024-01697-4}{{\em Commun. Phys.}
  {\bfseries 7} no.~1, (2024) 208},
  \href{http://arxiv.org/abs/2401.14570}{{\ttfamily arXiv:2401.14570
  [hep-lat]}}.

\bibitem{Than:2024zaj}
A.~T. Than, Y.~Y. Atas, A.~Chakraborty, J.~Zhang, M.~T. Diaz, K.~Wen, X.~Liu,
  R.~Lewis, A.~M. Green, C.~A. Muschik, and N.~M. Linke, ``The phase diagram of
  quantum chromodynamics in one dimension on a quantum computer,''
  \href{http://arxiv.org/abs/2501.00579}{{\ttfamily arXiv:2501.00579
  [quant-ph]}}. \url{https://arxiv.org/abs/2501.00579}.

\bibitem{Angelides2025first}
T.~Angelides, P.~Naredi, A.~Crippa, K.~Jansen, S.~K{\"u}hn, I.~Tavernelli, and
  D.~S. Wang, ``First-order phase transition of the schwinger model with a
  quantum computer,'' \href{http://dx.doi.org/10.1038/s41534-024-00950-6}{{\em
  npj Quantum Information} {\bfseries 11} no.~1, (2025) 6}.
  \url{https://doi.org/10.1038/s41534-024-00950-6}.

\bibitem{alexandrou2025realizingstringbreakingdynamics}
C.~Alexandrou, A.~Athenodorou, K.~Blekos, G.~Polykratis, and S.~Kühn,
  ``Realizing string breaking dynamics in a $z_2$ lattice gauge theory on
  quantum hardware,'' \href{http://arxiv.org/abs/2504.13760}{{\ttfamily
  arXiv:2504.13760 [hep-lat]}}. \url{https://arxiv.org/abs/2504.13760}.

\bibitem{Cochran:2024rwe}
T.~A. Cochran {\em et~al.}, ``{Visualizing dynamics of charges and strings in
  (2 + 1)D lattice gauge theories},''
  \href{http://dx.doi.org/10.1038/s41586-025-08999-9}{{\em Nature} {\bfseries
  642} no.~8067, (2025) 315--320},
  \href{http://arxiv.org/abs/2409.17142}{{\ttfamily arXiv:2409.17142
  [quant-ph]}}.

\bibitem{Gyawali:2024hrz}
G.~Gyawali {\em et~al.}, ``{Observation of disorder-free localization and
  efficient disorder averaging on a quantum processor},''
  \href{http://arxiv.org/abs/2410.06557}{{\ttfamily arXiv:2410.06557
  [quant-ph]}}.

\bibitem{gonzalezcuadra2024observationstringbreaking2}
D.~Gonz{\'a}lez-Cuadra, M.~Hamdan, T.~V. Zache, B.~Braverman, M.~Kornja{\v c}a,
  A.~Lukin, S.~H. Cant{\'u}, F.~Liu, S.-T. Wang, A.~Keesling, M.~D. Lukin,
  P.~Zoller, and A.~Bylinskii, ``Observation of string breaking on a (2 + 1)d
  rydberg quantum simulator,''
  \href{http://dx.doi.org/10.1038/s41586-025-09051-6}{{\em Nature} {\bfseries
  642} no.~8067, (2025) 321--326}.
  \url{https://doi.org/10.1038/s41586-025-09051-6}.

\bibitem{crippa2024analysisconfinementstring2}
A.~Crippa, K.~Jansen, and E.~Rinaldi, ``Analysis of the confinement string in
  (2 + 1)-dimensional quantum electrodynamics with a trapped-ion quantum
  computer,'' \href{http://arxiv.org/abs/2411.05628}{{\ttfamily
  arXiv:2411.05628 [hep-lat]}}. \url{https://arxiv.org/abs/2411.05628}.

\bibitem{schuhmacher2025observationhadronscatteringlattice}
J.~Schuhmacher, G.-X. Su, J.~J. Osborne, A.~Gandon, J.~C. Halimeh, and
  I.~Tavernelli, ``Observation of hadron scattering in a lattice gauge theory
  on a quantum computer,'' \href{http://arxiv.org/abs/2505.20387}{{\ttfamily
  arXiv:2505.20387 [quant-ph]}}. \url{https://arxiv.org/abs/2505.20387}.

\bibitem{davoudi2025quantumcomputationhadronscattering}
Z.~Davoudi, C.-C. Hsieh, and S.~V. Kadam, ``Quantum computation of hadron
  scattering in a lattice gauge theory,''
  \href{http://arxiv.org/abs/2505.20408}{{\ttfamily arXiv:2505.20408
  [quant-ph]}}. \url{https://arxiv.org/abs/2505.20408}.

\bibitem{Bernien2017}
H.~Bernien, S.~Schwartz, A.~Keesling, H.~Levine, A.~Omran, H.~Pichler, S.~Choi,
  A.~S. Zibrov, M.~Endres, M.~Greiner, V.~Vuleti{\'c}, and M.~D. Lukin,
  ``Probing many-body dynamics on a 51-atom quantum simulator,''
  \href{http://dx.doi.org/10.1038/nature24622}{{\em Nature} {\bfseries 551}
  no.~7682, (2017) 579--584}. \url{https://doi.org/10.1038/nature24622}.

\bibitem{Su2022}
G.-X. Su, H.~Sun, A.~Hudomal, J.-Y. Desaules, Z.-Y. Zhou, B.~Yang, J.~C.
  Halimeh, Z.-S. Yuan, Z.~Papi\ifmmode~\acute{c}\else \'{c}\fi{}, and J.-W.
  Pan, ``Observation of many-body scarring in a bose-hubbard quantum
  simulator,'' \href{http://dx.doi.org/10.1103/PhysRevResearch.5.023010}{{\em
  Phys. Rev. Res.} {\bfseries 5} (Apr, 2023) 023010}.
  \url{https://link.aps.org/doi/10.1103/PhysRevResearch.5.023010}.

\bibitem{Surace2020}
F.~M. Surace, P.~P. Mazza, G.~Giudici, A.~Lerose, A.~Gambassi, and M.~Dalmonte,
  ``Lattice gauge theories and string dynamics in {Rydberg} atom quantum
  simulators,'' \href{http://dx.doi.org/10.1103/PhysRevX.10.021041}{{\em Phys.
  Rev. X} {\bfseries 10} (May, 2020) 021041}.
  \url{https://link.aps.org/doi/10.1103/PhysRevX.10.021041}.

\bibitem{Desaules2021}
J.-Y. Desaules, K.~Bull, A.~Daniel, and Z.~Papi{\'c}, ``Hypergrid subgraphs and
  the origin of scarred quantum walks in the many-body {Hilbert} space,'' {\em
  arXiv preprint} (2021) , \href{http://arxiv.org/abs/2112.06885}{{\ttfamily
  arXiv:2112.06885}}.

\bibitem{Desaules2022prominent}
J.-Y. Desaules, A.~Hudomal, D.~Banerjee, A.~Sen, Z.~Papi\ifmmode~\acute{c}\else
  \'{c}\fi{}, and J.~C. Halimeh, ``Prominent quantum many-body scars in a
  truncated schwinger model,''
  \href{http://dx.doi.org/10.1103/PhysRevB.107.205112}{{\em Phys. Rev. B}
  {\bfseries 107} (May, 2023) 205112}.
  \url{https://link.aps.org/doi/10.1103/PhysRevB.107.205112}.

\bibitem{Banerjee2021}
D.~Banerjee and A.~Sen, ``Quantum scars from zero modes in an abelian lattice
  gauge theory on ladders,''
  \href{http://dx.doi.org/10.1103/PhysRevLett.126.220601}{{\em Phys. Rev.
  Lett.} {\bfseries 126} (Jun, 2021) 220601}.
  \url{https://link.aps.org/doi/10.1103/PhysRevLett.126.220601}.

\bibitem{Halimeh2022robust}
J.~C. Halimeh, L.~Barbiero, P.~Hauke, F.~Grusdt, and A.~Bohrdt, ``Robust
  quantum many-body scars in lattice gauge theories,''
  \href{http://dx.doi.org/10.22331/q-2023-05-15-1004}{{\em {Quantum}}
  {\bfseries 7} (May, 2023) 1004}.
  \url{https://doi.org/10.22331/q-2023-05-15-1004}.

\bibitem{biswas2022scars}
S.~Biswas, D.~Banerjee, and A.~Sen, ``{Scars from protected zero modes and
  beyond in $U(1)$ quantum link and quantum dimer models},''
  \href{http://dx.doi.org/10.21468/SciPostPhys.12.5.148}{{\em SciPost Phys.}
  {\bfseries 12} (2022) 148}.
  \url{https://scipost.org/10.21468/SciPostPhys.12.5.148}.

\bibitem{Sau:2023clm}
I.~Sau, P.~Stornati, D.~Banerjee, and A.~Sen, ``{Sublattice scars and beyond in
  two-dimensional U(1) quantum link lattice gauge theories},''
  \href{http://dx.doi.org/10.1103/PhysRevD.109.034519}{{\em Phys. Rev. D}
  {\bfseries 109} no.~3, (2024) 034519},
  \href{http://arxiv.org/abs/2311.06773}{{\ttfamily arXiv:2311.06773
  [hep-lat]}}.

\bibitem{osborne2024quantummanybodyscarring21d}
J.~Osborne, I.~P. McCulloch, and J.~C. Halimeh, ``Quantum many-body scarring in
  $2+1$d gauge theories with dynamical matter,''
  \href{http://arxiv.org/abs/2403.08858}{{\ttfamily arXiv:2403.08858
  [cond-mat.quant-gas]}}. \url{https://arxiv.org/abs/2403.08858}.

\bibitem{Budde2024qmbs}
T.~Budde, M.~Krstic~Marinkovic, and J.~C. Pinto~Barros, ``Quantum many-body
  scars for arbitrary integer spin in $2+1\mathrm{D}$ abelian gauge theories,''
  \href{http://dx.doi.org/10.1103/PhysRevD.110.094506}{{\em Phys. Rev. D}
  {\bfseries 110} (Nov, 2024) 094506}.
  \url{https://link.aps.org/doi/10.1103/PhysRevD.110.094506}.

\bibitem{Calajo2025QMBS}
G.~Calaj\'o, G.~Cataldi, M.~Rigobello, D.~Wanisch, G.~Magnifico, P.~Silvi,
  S.~Montangero, and J.~C. Halimeh, ``Quantum many-body scarring in a
  non-abelian lattice gauge theory,''
  \href{http://dx.doi.org/10.1103/PhysRevResearch.7.013322}{{\em Phys. Rev.
  Res.} {\bfseries 7} (Mar, 2025) 013322}.
  \url{https://link.aps.org/doi/10.1103/PhysRevResearch.7.013322}.

\bibitem{Sala2020ergodicity}
P.~Sala, T.~Rakovszky, R.~Verresen, M.~Knap, and F.~Pollmann, ``Ergodicity
  breaking arising from hilbert space fragmentation in dipole-conserving
  hamiltonians,'' \href{http://dx.doi.org/10.1103/PhysRevX.10.011047}{{\em
  Phys. Rev. X} {\bfseries 10} (Feb, 2020) 011047}.
  \url{https://link.aps.org/doi/10.1103/PhysRevX.10.011047}.

\bibitem{Khemani2020localization}
V.~Khemani, M.~Hermele, and R.~Nandkishore, ``Localization from hilbert space
  shattering: From theory to physical realizations,''
  \href{http://dx.doi.org/10.1103/PhysRevB.101.174204}{{\em Phys. Rev. B}
  {\bfseries 101} (May, 2020) 174204}.
  \url{https://link.aps.org/doi/10.1103/PhysRevB.101.174204}.

\bibitem{Jeyaretnam2025Hilbert}
J.~Jeyaretnam, T.~Bhore, J.~J. Osborne, J.~C. Halimeh, and Z.~Papi{\'c},
  ``Hilbert space fragmentation at the origin of disorder-free localization in
  the lattice schwinger model,''
  \href{http://dx.doi.org/10.1038/s42005-025-02039-8}{{\em Communications
  Physics} {\bfseries 8} no.~1, (2025) 172}.
  \url{https://doi.org/10.1038/s42005-025-02039-8}.

\bibitem{ciavarella2025generichilbertspacefragmentation}
A.~N. Ciavarella, C.~W. Bauer, and J.~C. Halimeh, ``Generic hilbert space
  fragmentation in kogut--susskind lattice gauge theories,''
  \href{http://arxiv.org/abs/2502.03533}{{\ttfamily arXiv:2502.03533
  [quant-ph]}}. \url{https://arxiv.org/abs/2502.03533}.

\bibitem{datla2025statisticallocalizationrydbergsimulator}
P.~R. Datla, L.~Zhao, W.~W. Ho, N.~Klco, and H.~Loh, ``Statistical localization
  in a rydberg simulator of $u(1)$ lattice gauge theory,''
  \href{http://arxiv.org/abs/2505.18143}{{\ttfamily arXiv:2505.18143
  [quant-ph]}}. \url{https://arxiv.org/abs/2505.18143}.

\bibitem{Smith2017}
A.~Smith, J.~Knolle, D.~L. Kovrizhin, and R.~Moessner, ``Disorder-free
  localization,'' \href{http://dx.doi.org/10.1103/PhysRevLett.118.266601}{{\em
  Phys. Rev. Lett.} {\bfseries 118} (Jun, 2017) 266601}.
  \url{https://link.aps.org/doi/10.1103/PhysRevLett.118.266601}.

\bibitem{Brenes2018}
M.~Brenes, M.~Dalmonte, M.~Heyl, and A.~Scardicchio, ``Many-body localization
  dynamics from gauge invariance,''
  \href{http://dx.doi.org/10.1103/PhysRevLett.120.030601}{{\em Phys. Rev.
  Lett.} {\bfseries 120} (Jan, 2018) 030601}.
  \url{https://link.aps.org/doi/10.1103/PhysRevLett.120.030601}.

\bibitem{Muschik2017}
C.~Muschik, M.~Heyl, E.~Martinez, T.~Monz, P.~Schindler, B.~Vogell,
  M.~Dalmonte, P.~Hauke, R.~Blatt, and P.~Zoller, ``U(1) wilson lattice gauge
  theories in digital quantum simulators,''
  \href{http://dx.doi.org/10.1088/1367-2630/aa89ab}{{\em New J. Phys.}
  {\bfseries 19} no.~10, (Oct, 2017) 103020}.
  \url{https://doi.org/10.1088/1367-2630/aa89ab}.

\bibitem{Joshi2025efficient}
Joshi et al., ``efficient qudit circuit for quench dynamics of $2 + 1$d quantum
  link electrodynamics'' (in preparation, 2025).

\bibitem{Byrnes:2005qx}
T.~Byrnes and Y.~Yamamoto, ``{Simulating lattice gauge theories on a quantum
  computer},'' \href{http://dx.doi.org/10.1103/PhysRevA.73.022328}{{\em Phys.
  Rev. A} {\bfseries 73} (2006) 022328},
  \href{http://arxiv.org/abs/quant-ph/0510027}{{\ttfamily
  arXiv:quant-ph/0510027}}.

\bibitem{Klco2020}
N.~Klco, M.~J. Savage, and J.~R. Stryker, ``{SU(2)} {non-Abelian} gauge field
  theory in one dimension on digital quantum computers,''
  \href{http://dx.doi.org/10.1103/PhysRevD.101.074512}{{\em Phys. Rev. D}
  {\bfseries 101} (Apr, 2020) 074512}.
  \url{https://link.aps.org/doi/10.1103/PhysRevD.101.074512}.

\bibitem{Raychowdhury2020loop}
I.~Raychowdhury and J.~R. Stryker, ``Loop, string, and hadron dynamics in su(2)
  hamiltonian lattice gauge theories,''
  \href{http://dx.doi.org/10.1103/PhysRevD.101.114502}{{\em Phys. Rev. D}
  {\bfseries 101} (Jun, 2020) 114502}.
  \url{https://link.aps.org/doi/10.1103/PhysRevD.101.114502}.

\bibitem{Halimeh2023Spin}
J.~C. Halimeh, L.~Homeier, A.~Bohrdt, and F.~Grusdt, ``{Spin Exchange-Enabled
  Quantum Simulator for Large-Scale Non-Abelian Gauge Theories},''
  \href{http://dx.doi.org/10.1103/PRXQuantum.5.030358}{{\em PRX Quantum}
  {\bfseries 5} no.~3, (2024) 030358},
  \href{http://arxiv.org/abs/2305.06373}{{\ttfamily arXiv:2305.06373
  [cond-mat.quant-gas]}}.

\bibitem{Surace2023scalable}
F.~M. Surace, P.~Fromholz, F.~Scazza, and M.~Dalmonte, ``Scalable, ab initio
  protocol for quantum simulating {SU}({$N$}){$\times$}u(1) {L}attice {G}auge
  {T}heories,'' \href{http://dx.doi.org/10.22331/q-2024-05-23-1359}{{\em
  {Quantum}} {\bfseries 8} (May, 2024) 1359}.
  \url{https://doi.org/10.22331/q-2024-05-23-1359}.

\bibitem{Fontana:2024rux}
P.~Fontana, M.~M. Riaza, and A.~Celi, ``{An efficient finite-resource
  formulation of non-Abelian lattice gauge theories beyond one dimension},''
  \href{http://arxiv.org/abs/2409.04441}{{\ttfamily arXiv:2409.04441
  [quant-ph]}}.

\bibitem{illa2025improvedhoneycombhyperhoneycomblattice}
M.~Illa, M.~J. Savage, and X.~Yao, ``Improved honeycomb and hyper-honeycomb
  lattice hamiltonians for quantum simulations of non-abelian gauge theories,''
  \href{http://arxiv.org/abs/2503.09688}{{\ttfamily arXiv:2503.09688
  [hep-lat]}}. \url{https://arxiv.org/abs/2503.09688}.

\bibitem{depaciani2025quantumsimulationfermionicnonabelian}
G.~D. Paciani, L.~Homeier, J.~C. Halimeh, M.~Aidelsburger, and F.~Grusdt,
  ``Quantum simulation of fermionic non-abelian lattice gauge theories in
  $(2+1)$d with built-in gauge protection,''
  \href{http://arxiv.org/abs/2506.14747}{{\ttfamily arXiv:2506.14747
  [cond-mat.quant-gas]}}. \url{https://arxiv.org/abs/2506.14747}.

\bibitem{Hanada:2025yzx}
M.~Hanada, S.~Matsuura, E.~Mendicelli, and E.~Rinaldi, ``{Exponential
  improvement in quantum simulations of bosons},''
  \href{http://arxiv.org/abs/2505.02553}{{\ttfamily arXiv:2505.02553
  [quant-ph]}}.

\bibitem{bergner2025exponentialspeedupquantumsimulation}
G.~Bergner and M.~Hanada, ``Exponential speedup in quantum simulation of
  kogut-susskind hamiltonian via orbifold lattice,''
  \href{http://arxiv.org/abs/2506.00755}{{\ttfamily arXiv:2506.00755
  [quant-ph]}}. \url{https://arxiv.org/abs/2506.00755}.

\bibitem{Buser:2020cvn}
A.~J. Buser, H.~Gharibyan, M.~Hanada, M.~Honda, and J.~Liu, ``{Quantum
  simulation of gauge theory via orbifold lattice},''
  \href{http://dx.doi.org/10.1007/JHEP09(2021)034}{{\em JHEP} {\bfseries 09}
  (2021) 034}, \href{http://arxiv.org/abs/2011.06576}{{\ttfamily
  arXiv:2011.06576 [hep-th]}}.

\bibitem{Bergner:2024qjl}
G.~Bergner, M.~Hanada, E.~Rinaldi, and A.~Schafer, ``{Toward QCD on quantum
  computer: orbifold lattice approach},''
  \href{http://dx.doi.org/10.1007/JHEP05(2024)234}{{\em JHEP} {\bfseries 05}
  (2024) 234}, \href{http://arxiv.org/abs/2401.12045}{{\ttfamily
  arXiv:2401.12045 [hep-th]}}.

\bibitem{Kaplan:2002wv}
D.~B. Kaplan, E.~Katz, and M.~\"{U}nsal, ``{Supersymmetry on a spatial
  lattice},'' \href{http://dx.doi.org/10.1088/1126-6708/2003/05/037}{{\em JHEP}
  {\bfseries 05} (2003) 037},
  \href{http://arxiv.org/abs/hep-lat/0206019}{{\ttfamily
  arXiv:hep-lat/0206019}}.

\bibitem{Rhodes:2024zbr}
M.~L. Rhodes, M.~Kreshchuk, and S.~Pathak, ``{Exponential Improvements in the
  Simulation of Lattice Gauge Theories Using Near-Optimal Techniques},''
  \href{http://dx.doi.org/10.1103/PRXQuantum.5.040347}{{\em PRX Quantum}
  {\bfseries 5} no.~4, (2024) 040347},
  \href{http://arxiv.org/abs/2405.10416}{{\ttfamily arXiv:2405.10416
  [quant-ph]}}.

\bibitem{srivatsa2025bosonicvsfermionicmatter}
N.~S. Srivatsa, J.~J. Osborne, D.~Banerjee, and J.~C. Halimeh, ``Bosonic vs.
  fermionic matter in quantum simulations of $2+1$d gauge theories,''
  \href{http://arxiv.org/abs/2504.17000}{{\ttfamily arXiv:2504.17000
  [cond-mat.str-el]}}. \url{https://arxiv.org/abs/2504.17000}.

\bibitem{Hanada:2022pps}
M.~Hanada, J.~Liu, E.~Rinaldi, and M.~Tezuka, ``{Estimating truncation effects
  of quantum bosonic systems using sampling algorithms},''
  \href{http://dx.doi.org/10.1088/2632-2153/ad035c}{{\em Mach. Learn. Sci.
  Tech.} {\bfseries 4} no.~4, (2023) 045021},
  \href{http://arxiv.org/abs/2212.08546}{{\ttfamily arXiv:2212.08546
  [quant-ph]}}.

\bibitem{Coppersmith:2002skh}
D.~Coppersmith, ``{An approximate Fourier transform useful in quantum
  factoring},'' \href{http://arxiv.org/abs/quant-ph/0201067}{{\ttfamily
  arXiv:quant-ph/0201067}}.

\bibitem{Campbell:2020wqh}
E.~T. Campbell, ``{Early fault-tolerant simulations of the Hubbard model},''
  \href{http://dx.doi.org/10.1088/2058-9565/ac3110}{{\em Quantum Sci. Technol.}
  {\bfseries 7} no.~1, (2022) 015007},
  \href{http://arxiv.org/abs/2012.09238}{{\ttfamily arXiv:2012.09238
  [quant-ph]}}.

\bibitem{Jordan_Wigner}
P.~{Jordan} and E.~{Wigner}, ``{{\"U}ber das Paulische {\"A}quivalenzverbot},''
  \href{http://dx.doi.org/10.1007/BF01331938}{{\em Zeitschrift fur Physik}
  {\bfseries 47} no.~9-10, (Sept., 1928) 631--651}.

\bibitem{Bravyi2002AnPhy}
S.~B. {Bravyi} and A.~Y. {Kitaev}, ``{Fermionic Quantum Computation},''
  \href{http://dx.doi.org/10.1006/aphy.2002.6254}{{\em Annals of Physics}
  {\bfseries 298} no.~1, (May, 2002) 210--226},
  \href{http://arxiv.org/abs/quant-ph/0003137}{{\ttfamily
  arXiv:quant-ph/0003137 [quant-ph]}}.

\bibitem{Yu-An2018AnPhy}
Y.-A. Chen and A.~Kapustin, ``Bosonization in three spatial dimensions and a
  2-form gauge theory,''
  \href{http://dx.doi.org/10.1103/PhysRevB.100.245127}{{\em Phys. Rev. B}
  {\bfseries 100} (Dec, 2019) 245127}.
  \url{https://link.aps.org/doi/10.1103/PhysRevB.100.245127}.

\bibitem{Verstraete-Cirac2005}
F.~{Verstraete} and J.~I. {Cirac}, ``{Mapping local Hamiltonians of fermions to
  local Hamiltonians of spins},''
  \href{http://dx.doi.org/10.1088/1742-5468/2005/09/P09012}{{\em Journal of
  Statistical Mechanics: Theory and Experiment} {\bfseries 2005} no.~9, (Sept.,
  2005) 09012}, \href{http://arxiv.org/abs/cond-mat/0508353}{{\ttfamily
  arXiv:cond-mat/0508353 [cond-mat.str-el]}}.

\bibitem{Kitaev2006AnPhy}
A.~{Kitaev}, ``{Anyons in an exactly solved model and beyond},''
  \href{http://dx.doi.org/10.1016/j.aop.2005.10.005}{{\em Annals of Physics}
  {\bfseries 321} no.~1, (Jan., 2006) 2--111},
  \href{http://arxiv.org/abs/cond-mat/0506438}{{\ttfamily
  arXiv:cond-mat/0506438 [cond-mat.mes-hall]}}.

\bibitem{Whitfield_2016}
J.~D. Whitfield, V.~Havlíček, and M.~Troyer, ``Local spin operators for
  fermion simulations,''
  \href{http://dx.doi.org/10.1103/physreva.94.030301}{{\em Physical Review A}
  {\bfseries 94} no.~3, (Sept., 2016) }.
  \url{http://dx.doi.org/10.1103/PhysRevA.94.030301}.

\bibitem{Jiang2019PhRvP}
Z.~{Jiang}, J.~{McClean}, R.~{Babbush}, and H.~{Neven}, ``{Majorana Loop
  Stabilizer Codes for Error Mitigation in Fermionic Quantum Simulations},''
  \href{http://dx.doi.org/10.1103/PhysRevApplied.12.064041}{{\em Physical
  Review Applied} {\bfseries 12} no.~6, (Dec., 2019) 064041},
  \href{http://arxiv.org/abs/1812.08190}{{\ttfamily arXiv:1812.08190
  [quant-ph]}}.

\bibitem{Setia2019PhRvR}
K.~{Setia}, S.~{Bravyi}, A.~{Mezzacapo}, and J.~D. {Whitfield}, ``{Superfast
  encodings for fermionic quantum simulation},''
  \href{http://dx.doi.org/10.1103/PhysRevResearch.1.033033}{{\em Physical
  Review Research} {\bfseries 1} no.~3, (Oct., 2019) 033033},
  \href{http://arxiv.org/abs/1810.05274}{{\ttfamily arXiv:1810.05274
  [quant-ph]}}.

\bibitem{Bochniak2020JHEP}
A.~{Bochniak} and B.~{Ruba}, ``{Bosonization based on Clifford algebras and its
  gauge theoretic interpretation},''
  \href{http://dx.doi.org/10.1007/JHEP12(2020)118}{{\em Journal of High Energy
  Physics} {\bfseries 2020} no.~12, (Dec., 2020) 118},
  \href{http://arxiv.org/abs/2003.06905}{{\ttfamily arXiv:2003.06905
  [math-ph]}}.

\bibitem{Derby_2021}
C.~Derby, J.~Klassen, J.~Bausch, and T.~Cubitt, ``Compact fermion to qubit
  mappings,'' \href{http://dx.doi.org/10.1103/physrevb.104.035118}{{\em
  Physical Review B} {\bfseries 104} no.~3, (July, 2021) }.
  \url{http://dx.doi.org/10.1103/PhysRevB.104.035118}.

\bibitem{Po2021arXiv210710842P}
H.~C. {Po}, ``{Symmetric Jordan-Wigner transformation in higher dimensions},''
  \href{http://dx.doi.org/10.48550/arXiv.2107.10842}{{\em arXiv e-prints}
  (July, 2021) arXiv:2107.10842},
  \href{http://arxiv.org/abs/2107.10842}{{\ttfamily arXiv:2107.10842
  [cond-mat.str-el]}}.

\bibitem{Chen:2018nog}
Y.-A. Chen and A.~Kapustin, ``{Bosonization in three spatial dimensions and a
  2-form gauge theory},''
  \href{http://dx.doi.org/10.1103/PhysRevB.100.245127}{{\em Phys. Rev. B}
  {\bfseries 100} no.~24, (2019) 245127},
  \href{http://arxiv.org/abs/1807.07081}{{\ttfamily arXiv:1807.07081
  [cond-mat.str-el]}}.

\bibitem{Chen2020PhRvR}
Y.-A. {Chen}, ``{Exact bosonization in arbitrary dimensions},''
  \href{http://dx.doi.org/10.1103/PhysRevResearch.2.033527}{{\em Physical
  Review Research} {\bfseries 2} no.~3, (Sept., 2020) 033527},
  \href{http://arxiv.org/abs/1911.00017}{{\ttfamily arXiv:1911.00017
  [cond-mat.str-el]}}.

\bibitem{Chen2023PRXQ}
Y.-A. Chen and Y.~Xu, ``{Equivalence between Fermion-to-Qubit Mappings in two
  Spatial Dimensions},''
  \href{http://dx.doi.org/10.1103/PRXQuantum.4.010326}{{\em PRX Quantum}
  {\bfseries 4} no.~1, (2023) 010326},
  \href{http://arxiv.org/abs/2201.05153}{{\ttfamily arXiv:2201.05153
  [quant-ph]}}.

\bibitem{Verstraete:2005pn}
F.~Verstraete and J.~I. Cirac, ``{Mapping local Hamiltonians of fermions to
  local Hamiltonians of spins},''
  \href{http://dx.doi.org/10.1088/1742-5468/2005/09/P09012}{{\em J. Stat.
  Mech.} {\bfseries 0509} (2005) P09012},
  \href{http://arxiv.org/abs/cond-mat/0508353}{{\ttfamily
  arXiv:cond-mat/0508353}}.

\bibitem{OBrien2018}
T.~E. O'Brien, P.~Ro\ifmmode~\dot{z}\else \.{z}\fi{}ek, and A.~R. Akhmerov,
  ``Majorana-based fermionic quantum computation,''
  \href{http://dx.doi.org/10.1103/PhysRevLett.120.220504}{{\em Phys. Rev.
  Lett.} {\bfseries 120} (Jun, 2018) 220504}.
  \url{https://link.aps.org/doi/10.1103/PhysRevLett.120.220504}.

\bibitem{Gonzalez2023}
D.~{Gonz{\'a}lez-Cuadra}, D.~{Bluvstein}, M.~{Kalinowski}, R.~{Kaubruegger},
  N.~{Maskara}, P.~{Naldesi}, T.~V. {Zache}, A.~M. {Kaufman}, M.~D. {Lukin},
  H.~{Pichler}, B.~{Vermersch}, J.~{Ye}, and P.~{Zoller}, ``{Fermionic quantum
  processing with programmable neutral atom arrays},''
  \href{http://dx.doi.org/10.1073/pnas.2304294120}{{\em Proceedings of the
  National Academy of Science} {\bfseries 120} no.~35, (Aug., 2023)
  e2304294120}, \href{http://arxiv.org/abs/2303.06985}{{\ttfamily
  arXiv:2303.06985 [quant-ph]}}.

\bibitem{Vilkelis2024}
K.~{Vilkelis}, A.~L.~R. {Manesco}, J.~D. {Torres Luna}, S.~{Miles},
  M.~{Wimmer}, and A.~R. {Akhmerov}, ``{Fermionic quantum computation with
  Cooper pair splitters},''
  \href{http://dx.doi.org/10.21468/SciPostPhys.16.5.135}{{\em SciPost Physics}
  {\bfseries 16} no.~5, (May, 2024) 135},
  \href{http://arxiv.org/abs/2309.00447}{{\ttfamily arXiv:2309.00447
  [cond-mat.mes-hall]}}.

\bibitem{Clark:2006fx}
M.~A. Clark and A.~D. Kennedy, ``{Accelerating dynamical fermion computations
  using the rational hybrid Monte Carlo (RHMC) algorithm with multiple
  pseudofermion fields},''
  \href{http://dx.doi.org/10.1103/PhysRevLett.98.051601}{{\em Phys. Rev. Lett.}
  {\bfseries 98} (2007) 051601},
  \href{http://arxiv.org/abs/hep-lat/0608015}{{\ttfamily
  arXiv:hep-lat/0608015}}.

\bibitem{hanada2022mcmc}
M.~Hanada and S.~Matsuura, {\em MCMC from scratch: A practical introduction to
  Markov chain Monte Carlo}.
\newblock Springer Nature, 2022.

\bibitem{Anagnostopoulos:2007fw}
K.~N. Anagnostopoulos, M.~Hanada, J.~Nishimura, and S.~Takeuchi, ``{Monte Carlo
  studies of supersymmetric matrix quantum mechanics with sixteen supercharges
  at finite temperature},''
  \href{http://dx.doi.org/10.1103/PhysRevLett.100.021601}{{\em Phys. Rev.
  Lett.} {\bfseries 100} (2008) 021601},
  \href{http://arxiv.org/abs/0707.4454}{{\ttfamily arXiv:0707.4454 [hep-th]}}.

\bibitem{Catterall:2007fp}
S.~Catterall and T.~Wiseman, ``{Towards lattice simulation of the gauge theory
  duals to black holes and hot strings},''
  \href{http://dx.doi.org/10.1088/1126-6708/2007/12/104}{{\em JHEP} {\bfseries
  12} (2007) 104}, \href{http://arxiv.org/abs/0706.3518}{{\ttfamily
  arXiv:0706.3518 [hep-lat]}}.

\bibitem{Gharibyan:2020bab}
H.~Gharibyan, M.~Hanada, M.~Honda, and J.~Liu, ``{Toward simulating
  superstring/M-theory on a quantum computer},''
  \href{http://dx.doi.org/10.1007/JHEP07(2021)140}{{\em JHEP} {\bfseries 07}
  (2021) 140}, \href{http://arxiv.org/abs/2011.06573}{{\ttfamily
  arXiv:2011.06573 [hep-th]}}.

\bibitem{Ishii:2008ib}
T.~Ishii, G.~Ishiki, S.~Shimasaki, and A.~Tsuchiya, ``{N=4 Super Yang-Mills
  from the Plane Wave Matrix Model},''
  \href{http://dx.doi.org/10.1103/PhysRevD.78.106001}{{\em Phys. Rev. D}
  {\bfseries 78} (2008) 106001},
  \href{http://arxiv.org/abs/0807.2352}{{\ttfamily arXiv:0807.2352 [hep-th]}}.

\bibitem{Hanada:2010kt}
M.~Hanada, S.~Matsuura, and F.~Sugino, ``{Two-dimensional lattice for
  four-dimensional N=4 supersymmetric Yang-Mills},''
  \href{http://dx.doi.org/10.1143/PTP.126.597}{{\em Prog. Theor. Phys.}
  {\bfseries 126} (2011) 597--611},
  \href{http://arxiv.org/abs/1004.5513}{{\ttfamily arXiv:1004.5513 [hep-lat]}}.

\bibitem{Hanada:2010gs}
M.~Hanada, ``{A proposal of a fine tuning free formulation of 4d N = 4 super
  Yang-Mills},'' \href{http://dx.doi.org/10.1007/JHEP11(2010)112}{{\em JHEP}
  {\bfseries 11} (2010) 112}, \href{http://arxiv.org/abs/1009.0901}{{\ttfamily
  arXiv:1009.0901 [hep-lat]}}.

\end{thebibliography}\endgroup

\end{document}